\documentclass[12pt]{article}

\pdfoutput=1

\usepackage[bulletsep]{collref}
\usepackage{amssymb,graphicx}
\usepackage[intlimits]{amsmath}
\usepackage{bbm}
\usepackage[small]{subfigure}

\usepackage{MnSymbol}

\makeatletter \@addtoreset{equation}{section} \makeatother

\makeatletter
\let\old@startsection=\@startsection
\let\oldl@section=\l@section
\renewcommand{\@startsection}[6]{\old@startsection{#1}{#2}{#3}{#4}{#5}{#6\mathversion{bold}}}
\renewcommand{\l@section}[2]{\oldl@section{\mathversion{bold}#1}{#2}}
\makeatother

\makeatletter
\let\old@makecaption=\@makecaption
\def\@makecaption{\small\old@makecaption}
\makeatother

\begin{document}

\newcommand{\be}{\begin{equation}}\newcommand{\ee}{\end{equation}}
\newcommand{\bea}{\begin{eqnarray}} \newcommand{\eea}{\end{eqnarray}}
\def\p{\partial}
\def\pa{\partial}
\def\ov{\over }
\def\a{\alpha }
\def\g{\gamma}
\def\s{\sigma }
\def\td{\tilde }
\def\vp{\varphi}
\def\gd{\nu }
\def \ha {{1 \over 2}}

\def\KK{{\cal K}}

\def\Xint#1{\mathchoice
{\XXint\displaystyle\textstyle{#1}} 
{\XXint\textstyle\scriptstyle{#1}} 
{\XXint\scriptstyle\scriptscriptstyle{#1}} 
{\XXint\scriptscriptstyle\scriptscriptstyle{#1}} 
\!\int}
\def\XXint#1#2#3{{\setbox0=\hbox{$#1{#2#3}{\int}$ }
\vcenter{\hbox{$#2#3$ }}\kern-.5\wd0}}
\def\ddashint{\Xint=}
\def\dashint{\Xint-}

\newcommand\cev[1]{\overleftarrow{#1}} 

\begin{flushright}\footnotesize
\texttt{NORDITA-2013-65} \\
\texttt{UUITP-11/13}
\vspace{0.6cm}
\end{flushright}

\renewcommand{\thefootnote}{\fnsymbol{footnote}}
\setcounter{footnote}{0}

\begin{center}
{\Large\textbf{\mathversion{bold} Massive ${\cal N}=2$  Gauge  Theories at Large $N$}
\par}

\vspace{0.8cm}

\textrm{J.G.~Russo$^{1,2}$ and
K.~Zarembo$^{3,4}$\footnote{Also at ITEP, Moscow, Russia.}}
\vspace{4mm}

\textit{${}^1$ Instituci\'o Catalana de Recerca i Estudis Avan\c cats (ICREA), \\
Pg. Lluis Companys, 23, 08010 Barcelona, Spain}\\
\textit{${}^2$  Department ECM, Institut de Ci\`encies del Cosmos,  \\
Universitat de Barcelona, Mart\'\i \ Franqu\`es, 1, 08028 Barcelona, Spain}\\
\textit{${}^3$Nordita, KTH Royal Institute of Technology and Stockholm University,
Roslagstullsbacken 23, SE-106 91 Stockholm, Sweden}\\
\textit{${}^4$Department of Physics and Astronomy, Uppsala University\\
SE-751 08 Uppsala, Sweden}\\
\vspace{0.2cm}
\texttt{jorge.russo@icrea.cat, zarembo@nordita.org}

\vspace{3mm}


\par\vspace{1cm}

\textbf{Abstract} \vspace{3mm}

\begin{minipage}{13cm}

Using exact results obtained from localization on $S^4$, we explore the large $N$ limit of ${\cal N}=2$ 
super Yang-Mills   theories with massive matter multiplets.
We focus on three cases: ${\cal N}=2^*$ theory, describing a massive hypermultiplet in  the adjoint representation, $SU(N)$ super-Yang-Mills with $2N$ massive hypermultiplets in the fundamental, and super QCD  with  massive quarks. When the radius of the four-sphere is sent to infinity the theories at hand are described by
solvable matrix models, which exhibit a number of interesting phenomena including
quantum phase transitions at finite 't~Hooft coupling.

\end{minipage}

\end{center}

\vspace{0.5cm}



\setcounter{page}{1}
\renewcommand{\thefootnote}{\arabic{footnote}}
\setcounter{footnote}{0}

\section{Introduction}

In this paper, we investigate massive $\mathcal{N}=2$ supersymmetric gauge theories  in the multicolor limit by exploiting the results of supersymmetric localization. The path integral of any $\mathcal{N}=2$ theory on $S^4$ can be localized to a finite-dimensional matrix integral  \cite{Pestun:2007rz}, and our goal will be to study the resulting matrix models in the large-$N$ limit. 

The multicolor limit of non-Abelian $SU(N)$ gauge theories is known to simplify their dynamics without distorting essential features of the non-perturbative behavior. Moreover, the relationship to string theory, which is now believed to be inherent to any non-Abelian theory, becomes most transparent within the large-$N$ expansion \cite{Hooft:1973jz}. The cleanest manifestation of the large-$N$ gauge/string duality is the AdS/CFT correspondence \cite{Maldacena:1998re,Gubser:1998bc,Witten:1998qj}  where the large-$N$ limit corresponds to free, non-interacting strings in a curved space. The  dual field theory, $\mathcal{N}=4$ superconformal Yang-Mills, has been studied quite thoroughly in the planar approximation. Much less is known about less supersymmetric and non-conformal theories. 

As far as $\mathcal{N}=2$ theories are concerned, the standard approach is based on the Seiberg-Witten theory \cite{Seiberg:1994rs,Seiberg:1994aj}. The large-$N$ limit of the latter was investigated in \cite{Douglas:1995nw,Ferrari:2001mg} for pure gauge 
$\mathcal{N}=2$ super-Yang-Mills (SYM). 
The localization matrix models for various $\mathcal{N}=2$ theories were analyzed in the large-$N$ limit in \cite{Rey:2010ry,Passerini:2011fe,Bourgine:2011ie,Fraser:2011qa,Russo:2012kj,Russo:2012ay,Buchel:2013id,Russo:2013qaa}. 
One of the outcomes of this analysis is a direct verification of the gauge/string duality in the non-conformal setting of the mass-deformed $\mathcal{N}=2^*$ super-Yang-Mills (SYM) theory \cite{Buchel:2013id}. The large-$N$ vacuum structure of pure gauge $\mathcal{N}=2$ SYM \cite{Douglas:1995nw,Ferrari:2001mg} can be reproduced from localization as well \cite{Russo:2012ay}. 
Here we concentrate on various other massive theories, starting with $\mathcal{N}=2^*$ SYM.

An important  simplification of the multicolor limit is expected to arise in the non-perturbative  sector:  instanton contributions should become negligible at large-$N$ due to exponential suppression of the instanton weight. The instanton moduli integration may, in principle, overcome the exponential suppression thus leading to a large-$N$ phase transition \cite{Gross:1994mr}. We have searched for the instanton-induced phase transitions in a number of $\mathcal{N}=2$ theories (secs.~\ref{instantonsec1}, \ref{instantonsec2}), so far with negative results.

Instead, we have found a novel type of phase transitions, which take place in the infinite volume limit and are associated with appearance of new nearly massless particles in the spectrum. In the $\mathcal{N}=2^*$ theory, the phase transition of this kind separates the weak-coupling phase from the strong-coupling phases \cite{Russo:2013qaa}. As the 't~Hooft coupling is increased, the theory undergoes an infinite sequence of phase transitions with critical points accumulating at infinite coupling. The strongly coupled vacuum acquires a rather irregular, fractal structure at small scales in the field space. The agreement with the holographic description \cite{Buchel:2013id} is obtained after coarse-grained average over this small-scale structure. What implications such a fractal structure can have for the holographic duality is unclear to us. It is conceivable that the strong-coupling limit (beyond the leading order, discussed in \cite{Buchel:2013id}) is not unique and depends on how the infinite-volume limit is taken. As a first step towards a deeper understanding of these issues, we will perform a detailed study of the critical behavior near the first of these phase transitions.

It turns out that  quantum, weak/strong coupling phase transitions are  generic features of $\mathcal{N}=2$ theories with two mass scales or with a dimensionless coupling. For instance, we will find that super-QCD (SQCD) in the Veneziano limit \cite{Veneziano:1976wm} undergoes a third-order phase transition as the quark mass varies. In fact, the localization matrix model of $\mathcal{N}=2$ SQCD is much simpler, making a detailed analysis of the phase transition possible.

The weak-coupling expansion of massive $\mathcal{N}=2$ theories that we will study is also of some interest, as it illustrates some generic features of asymptotically free QFTs with two well separated scales: the dynamically generated scale $\Lambda $ and the ``kinematic" scale $M $. If $M\gg \Lambda $ perturbation theory should be a reasonable approximation, but only up to power-like corrections:
\begin{equation}\label{power-condensates}
 \mathcal{A}={\rm perturbative}+\sum_{n=1}^{\infty }C_n\left(\frac{\Lambda }{M}\right)^{2n}.
\end{equation}
The coefficients $C_n$ are in general  not calculable, unless the theory can be solved exactly, and, at best, can be parameterized by vacuum condensates, like in the ITEP sum rules. Using localization techniques it will be possible to compute the coefficients of the OPE expansion exactly in certain multicolor $\mathcal{N}=2$ theories. 

\section{Generalities}

The theories we are going to study will contain a single vector multiplet $(A_\mu , \psi^1_\alpha  , \psi^2_\alpha  , \Phi +i\Phi ')$ of the $SU(N)$ gauge symmetry, and a number of matter hypermultiplets $(\phi , \chi _\alpha ,\tilde{\chi }_\alpha ,\tilde{\phi })$ either in the adjoint or in the fundamental representation of $SU(N)$. Each hypermultiplet of mass $M$ should be accompanied by the $CPT$ conjugate of mass $-M$.
We also briefly discuss quiver-type theories with product gauge groups and bi-fundamental matter.

The $SU(N)$ gauge symmetry of $\mathcal{N}=2$ SYM theories is usually broken to $U(1)^{N-1}$ by the vev of the adjoint scalar in the vector multiplet:
\begin{equation}\label{Phivev}
 \left\langle \Phi \right\rangle=\mathop{\mathrm{diag}}\left(a_1,\ldots ,a_N\right).
\end{equation}
The supersymmetric localization reduces the path integral of the theory compactified on $S^4$ to a finite dimensional integral over the Coulomb moduli, the eigenvalues of the scalar vev \cite{Pestun:2007rz}\footnote{More precisely, the integral goes along a real section of the complex moduli space, see \cite{Pestun:2007rz} for a discussion.}:
\begin{equation}\label{Pestunpf}
 Z=\int_{}^{}Da\,\mathcal{Z}_{\rm 1-loop}(a)\left|\mathcal{Z}_{\rm inst}(a)\right|^2\,{\rm e}\,^{-S_{\rm cl}(a)},
\end{equation}
where $Da$ is the usual Vandermonde measure on Hermitean matrices:
\begin{equation}
 Da=\prod_{i}^{}da_i\,\delta \left(\sum_{i}^{}a_i\right)\prod_{i<j}^{}\left(a_i-a_j\right)^2.
\end{equation}
The classical action arises from the $\mathcal{R}\mathop{\mathrm{tr}}\Phi ^2/4$ coupling of the scalar to the curvature of $S^4$, which is necessary for maintaining supersymmetry (e.g. \cite{Festuccia:2011ws}), and is equal to
\begin{equation}
 S_{\rm cl}=\frac{8\pi ^2N}{\lambda }\sum_{i}^{}a_i^2,
\end{equation}
where $\lambda =g_{\rm YM}^2N$ is the 't~Hooft coupling. 

The one-loop factor $\mathcal{Z}_{\rm 1-loop}$ was computed in \cite{Pestun:2007rz} and is expressed in terms of a single function
\begin{equation}\label{functionH}
 H(x)\equiv \prod_{n=1}^\infty \left(1+\frac{x^2}{n^2}\right)^n \,{\rm e}\,^{-\frac{x^2}{n}} .
\end{equation}
Various properties of this function and of its logarithmic derivative
\begin{equation}\label{functionKK}
 \KK(x) \equiv  -\frac{H'(x)}{H(x)} =2x\sum_{n=1}^\infty \left(\frac{1}{n} -\frac{n}{n^2+x^2}\right)
\end{equation}
are listed in  appendix~\ref{kernelappendix}. The  various multiplets contribute  as follows:
\begin{eqnarray}
 {\rm Vector~multiplet:}&& \prod_{i<j}^{}H^2(a_i-a_j)
\nonumber \\
 {\rm Adjoint~hypermultiplet:}&& \prod_{i<j}^{}\frac{1}{H(a_i-a_j+M)H(a_i-a_j-M)}
\nonumber \\
 {\rm Fundamental~hypermultiplet:}&&\prod_{i}^{}\frac{1}{H(a_i+M)}\,.
\end{eqnarray}
The one-loop factor $\mathcal{Z}_{\rm 1-loop}$ is the product of these factors over the matter content of the theory. 

The instanton factor $\mathcal{Z}_{\rm inst}$ is the Nekrasov partition function \cite{Nekrasov:2002qd,Nekrasov:2003rj} with the equivariant parameters set equal: $\epsilon _1=\epsilon _2=1$. In the large-$N$ limit the instantons are suppressed and for the most part we will just drop the instanton factor, except for sections~\ref{instantonsec1}, \ref{instantonsec2} where we explicitly check that the one-instanton contribution is exponentially small at $N\rightarrow \infty $.  

Our conventions are such that the radius $R$ of the four-sphere is set to one. It is easy to recover the dependence on $R$, which we will occasionally do, by rescaling  all the dimensionful quantities by $R$: $a_i\rightarrow a_iR$ and $M\rightarrow MR$. This in particular means that the decompactification limit ($R\rightarrow \infty $) in the radius-one units corresponds to the infinite-mass limit $M\rightarrow \infty $. The equivariant parameters of the instanton partition function, equal to one when $R=1$, in arbitrary units are equal to $1/R$.

Apart from to the free energy,
\begin{equation}
 F=-\frac{1}{N^2}\,\ln Z,
\end{equation}
 localization also allows one to compute the expectation value of the circular Wilson loop which, in addition to the gauge field,  couples to the scalar from the vector multiplet:
\begin{equation}
 W(C)\equiv \left\langle \frac{1}{N}\,\mathop{\mathrm{tr}}{\rm P}\exp
 \left[\oint_C d\tau \,\left(i\dot{x}^\mu A_\mu + |\dot{x}|\Phi \right)\right]
 \right\rangle.
\end{equation}
If the contour $C$ runs along the big circle of the four-sphere, the path integral with the Wilson loop inserted still localizes to the matrix model. The localization amounts to replacing the fields by their classical values, $A_\mu =0$ and $\Phi $ given by (\ref{Phivev}), and subsequently integrating over the Coulomb moduli, so the Wilson loop expectation value maps to the exponential operator in the matrix model:
\begin{equation}
 W(C)=\left\langle \frac{1}{N}\sum_{i}^{}\,{\rm e}\,^{2\pi a_i}\right\rangle,
\end{equation}
where  the average is now defined  by the partition function  (\ref{Pestunpf}).

\section{$\mathcal{N}=2^*$  SYM}

There are two possible ways to view  this theory. One way is to start with $\mathcal{N}=4$ SYM, and add  specific dimension two and dimension three operators to the Lagrangian. The field content of $\mathcal{N}=4$ SYM, in the  $\mathcal{N}=2$ terms, consists of a vector multiplet and two massless adjoint hypermultiplets. The unique massive deformation that preserves half of the supersymmetry is  obtained by adding equal masses to the two hypermultiplets. The resulting theory constitutes the simplest relevant perturbation of the superconformal $\mathcal{N}=4$ theory away from the conformal point.  The difference between $\mathcal{N}=2^*$ SYM and $\mathcal{N}=4$ SYM disappears in the UV,  for instance on the four-sphere of a very small radius  $R\ll 1/M$. 

In the opposite limit, say for $MR\gg 1$, the hypermultiplets can be integrated out, leaving behind pure $\mathcal{N}=2$ SYM, to the leading approximation. As a relevant perturbation of a finite theory and due to $\mathcal{N}=2$ supersymmetry, $\mathcal{N}=2^*$ theory is also UV finite. Thus, one can alternatively view $\mathcal{N}=2^*$ theory as a convenient UV regularization of pure $\mathcal{N}=2$ SYM, with the hypermultiplet mass playing the r\^ole of  the UV cutoff and the finite 't~Hooft coupling playing the r\^ole of  the bare coupling. From that perspective, 
\begin{figure}[t]
\begin{center}
 \centerline{\includegraphics[width=3cm]{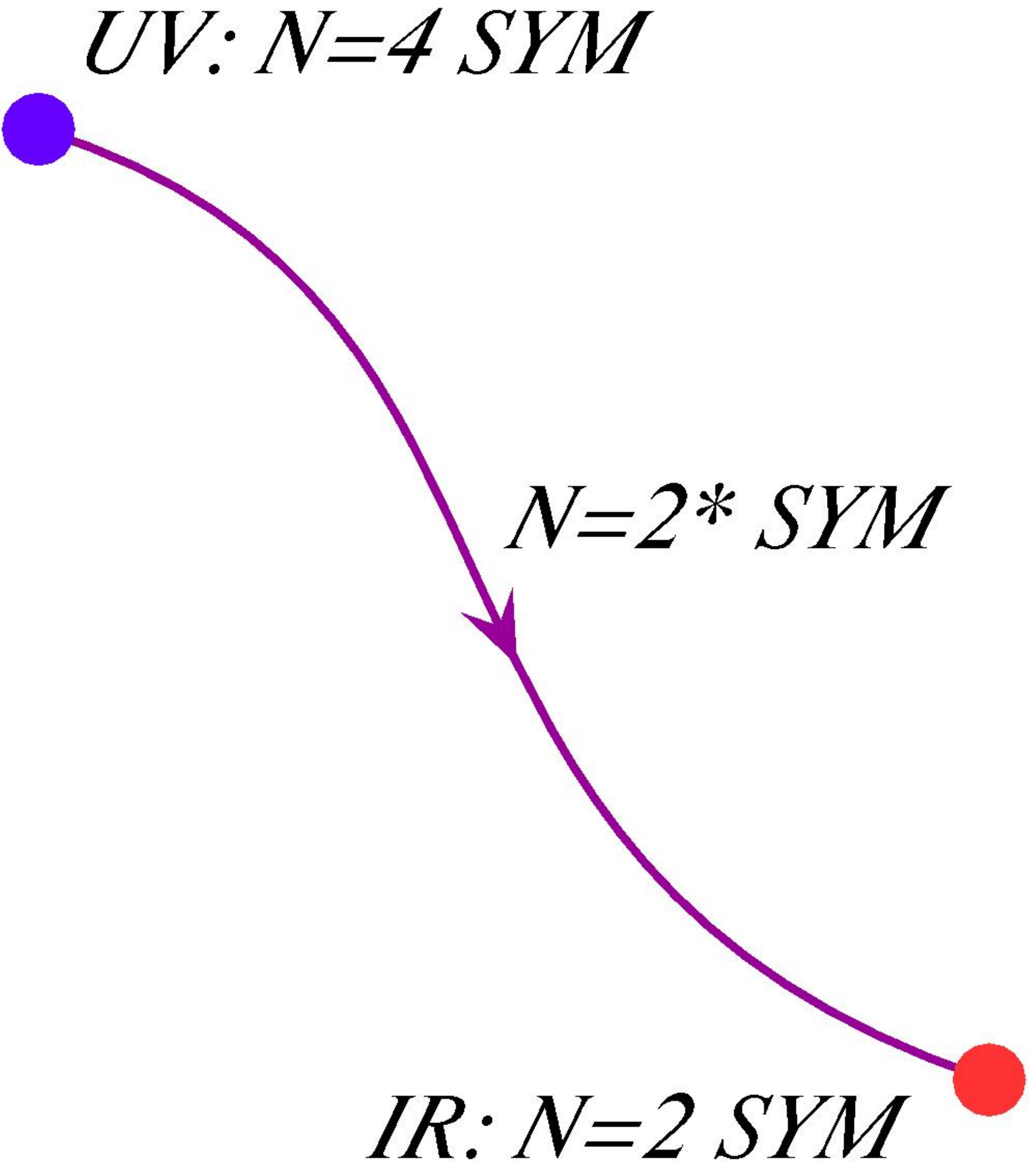}}
\caption{\label{N2*}\small $\mathcal{N}=2^*$ SYM as a flow from $\mathcal{N}=4$ SYM to $\mathcal{N}=2$ SYM.}
\end{center}
\end{figure}
$\mathcal{N}=2^*$ SYM describes a flow from $\mathcal{N}=4$ SYM in the UV to $\mathcal{N}=2$ SYM in the IR (fig.~\ref{N2*}). The hypermultiplet mass and the coupling constant combine into the dynamically generated scale of the $\mathcal{N}=2$ theory:
\begin{equation}\label{LambdaQCD}
 \Lambda =M\,{\rm e}\,^{-\frac{4\pi ^2}{\lambda }}.
\end{equation}
From this formula and (\ref{power-condensates}) we infer the general form of the weak-coupling expansion in the $\mathcal{N}=2^*$ SYM:
\begin{equation}\label{weak-expansion}
 \mathcal{A}={\rm perturbative}+\sum_{n=1}^{\infty }C_n\,{\rm e}\,^{-\frac{8\pi ^2n}{\lambda }}.
\end{equation}
This expansion can be interpreted as OPE, with the higher-order terms originating from irrelevant operators in the low-energy effective field theory obtained by integrating out the hypermultiplets \cite{Russo:2013qaa}.

It is important to realize that the flow picture only makes sense at weak coupling, when the scales $M$  and $\Lambda $ are well separated. When $\lambda $ is not very small, $M$ and $\Lambda $ are of the same order of magnitude and the effective field theory approximation  breaks down. Moreover, at strong coupling the suitably defined dynamical scale is parametrically larger than $M$. This follows from the supergravity analysis of the known holographic dual \cite{Pilch:2000ue} of $\mathcal{N}=2^*$ SYM \cite{Buchel:2000cn}, or can be derived directly from field theory using localization \cite{Buchel:2013id}.
The interactions at $\lambda \rightarrow \infty $ are so strong that the mass perturbation distorts  the dynamics at energy scales much bigger than $M$.

\subsection{Localization and large-$N$}

In order to use the localization results of \cite{Pestun:2007rz}, we compactify  $\mathcal{N}=2^*$ SYM on $S^4$ of radius $R$. We will be interested in the large-$N$ planar limit, in which the theory depends on two parameters, $\lambda $ and $MR$. Perhaps the most interesting regime is the decompactification limit $MR\rightarrow \infty $, but it is useful to keep $MR$ as an extra parameter, for instance in order to compare with $\mathcal{N}=4$ SYM at $MR\rightarrow 0$. 

\begin{figure}[t]
\begin{center}
 \centerline{\includegraphics[width=12cm]{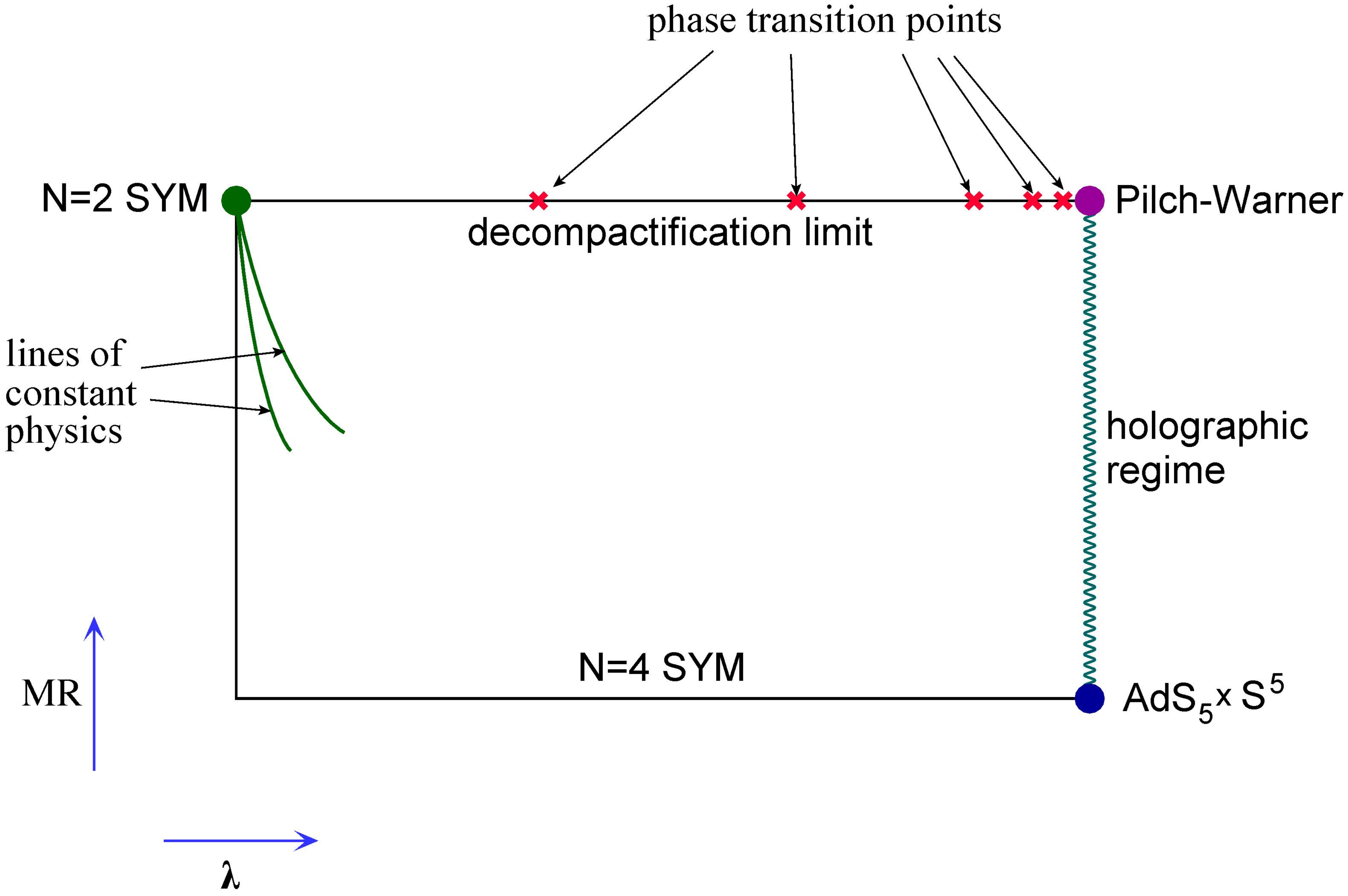}}
\caption{\label{phaseportrait}\small The phase diagram of $\mathcal{N}=2^*$ theory on $S^4$. On the sphere of a very small radius the theory is equivalent to $\mathcal{N}=4$ SYM. The opposite, decompactification limit corresponds to the theory defined on  flat $\mathbbm{R}^4$. Interpolation between $MR=0$ and $MR=\infty $ at weak coupling describes the RG flow from $\mathcal{N}=4$ SYM in the UV to $\mathcal{N}=2$ SYM in the IR.  The dynamical scale $\Lambda $ from (\ref{LambdaQCD}) takes on a particular fixed value on each line of constant physics. In the decompactification limit, the theory undergoes an infinite sequence of phase transitions that accumulate at strong coupling. The strong-coupling regime is supposed to have a weakly-curved holographic description at any $R$, but the dual geometry is known only in the two extreme cases: at $MR\rightarrow 0$ the theory becomes equivalent to $\mathcal{N}=4$ SYM, and the background should degenerate to $AdS_5\times S^5$ in the foliation where the boundary of $AdS_5$ is the round four-dimensional sphere; at $MR= \infty $, the dual background is the Pilch-Warner solution of the ten-dimensional type IIB supergravity \cite{Pilch:2000ue} that has a flat $\mathbbm{R}^4$ boundary.}
\end{center}
\end{figure}

The starting point of our analysis is the exact partition function \cite{Pestun:2007rz}:
\be\label{mint}
Z^{{\cal N}=2^*}=\int d^{N-1} a\, 
\prod _{i<j}\frac{(a_i-a_j)^2H^2(a_i-a_j)}{H(a_i-a_j-M)H(a_i-a_j+M)}
\,{\rm e}\,^{-\frac{8\pi^2N}{\lambda}\sum_i a_i^2 } \left|\mathcal{Z}_{\rm inst}\right|^2.
\ee
From now on we will set $\mathcal{Z}_{\rm inst}=1$, returning to instantons later to check if they are indeed exponentially suppressed at large $N$. The resulting matrix model, simple as it is, has unexpectedly rich dynamics. It captures the OPE expansion (\ref{weak-expansion}) at weak coupling \cite{Russo:2013qaa}, agrees with the string-theory predictions at strong coupling \cite{Buchel:2013id}, while at intermediate $\lambda $ it features an infinite sequence of quantum phase transitions, which only happen in the infinite-volume limit. The phase diagram of the model is shown in fig.~\ref{phaseportrait}.

In the planar limit, the matrix integral (\ref{mint})  is governed by
a saddle point. In terms of the eigenvalue density,
\begin{equation}
 \rho (x)=\frac{1}{N}\sum_{i}^{}\delta \left(x-a_i\right),
\end{equation}
the saddle-point equations are  equivalent to a singular integral equation:
\begin{equation}
 \label{nnstar}
\strokedint_{-\mu}^\mu dy \rho(y) 
\left(\frac{1}{x-y} -\KK(x-y)+\frac{1}{2}\,\KK(x-y+M)+\frac{1}{2}\,\KK(x-y-M)\right)= \frac{8\pi^2}{\lambda}\ x.
\end{equation}
The density $\rho (x)$ is defined on an interval $(-\mu,\mu)$ and is unit normalized. The integral equation only holds  for $x$ between $-\mu $ and $\mu $. Potentially possible two-cut solutions of the matrix model turn out to be  dynamically disfavored \cite{Russo:2012ay}. For clarity, we switched to the units where $R=1$. 

Once the saddle-point equation is solved, the circular Wilson loop is computed as
\begin{equation}
 W(C)=\int_{-\mu }^{\mu }dx\,\rho (x)\,{\rm e}\,^{2\pi x}\equiv 
 \left\langle \,{\rm e}\,^{2\pi x}\right\rangle.
\end{equation}
As far as the free energy is concerned, it is more convenient to calculate its first derivatives:
\begin{eqnarray}\label{dF}
 \frac{\partial F}{\partial \lambda }&=&-\frac{8\pi^2}{\lambda^2}\,
 \left\langle x^2\right\rangle
\nonumber \\
 \frac{\partial F}{\partial M}&=&\frac{1}{2}\left\langle \left\langle 
 \mathcal{K}(x-y-M)-\mathcal{K}(x-y+M)
 \right\rangle\right\rangle,
\end{eqnarray}
where the double brackets denote average over both $x$ and $y$.

We will analyze the saddle-point equation in detail at the extreme values of parameters, corresponding to the sides of the rectangle in fig.~\ref{phaseportrait}. We will also describe salient features of the solution for generic $M$ and $\lambda $, which can be obtained numerically. 

\subsection{Weak coupling}\label{N=2*weak:sec}

The weak-coupling limit of the $\mathcal{N}=2^*$ matrix model was studied in \cite{Russo:2012kj}. Here we add a few more remarks on the free energy and on the validity of the weak-coupling approximation.
At small $\lambda $, the classical force term on right-hand-side of the saddle-point equation squeezes the eigenvalue distribution towards zero making $\mu $ very small. We can then expand the kernel in powers of $x-y\sim \mu $. To the leading order we are left with just the Hilbert kernel. The solution to the integral equation is then described by the Wigner's semicircle:
\be\label{WignerLaw}
\rho(x)= \frac{2}{\pi\mu^2}\,\sqrt{\mu^2-x^2}\,,
\ee
with the width of the eigenvalue distribution given by $\mu =\sqrt{\lambda }/2\pi $.

The next order in the expansion can be taken into account without doing any new calculation. Indeed, the next term is linear in $x-y$, the $y$ term integrates to zero due to the $SU(N)$ constraint, and the $x$ term renormalizes the coupling constant on the right-hand-side:
\begin{equation}\label{lambdaR}
 \frac{8\pi ^2}{\lambda }~\longrightarrow~ \frac{8\pi ^2}{\lambda_R }\equiv 
 \frac{8\pi ^2}{\lambda }-\mathcal{K}'(M).
\end{equation}
At large $M$, we can use (\ref{KKapprox}) to approximate $\mathcal{K}(M)$. The effective coupling $\lambda _R$ then coincides with the running $\mathcal{N}=2$ Yang-Mills coupling renormalized at the scale set by the radius of the four-sphere:
\begin{equation}\label{renormalizedl}
 \frac{4 \pi ^2}{\lambda _R}\simeq \frac{4\pi ^2}{\lambda }-\ln M-1-\gamma
 =-\ln\Lambda -1-\gamma \qquad \left(M\gg 1\right).
\end{equation}
The lines of constant $\lambda _R$ are the lines of constant physics in the $\mathcal{N}=2$ theory.

The solution of the saddle-point equations is again the semicircle with the width determined by the renormalized coupling:
\begin{equation}\label{mmaa}
 \mu =\frac{\sqrt{\lambda _R}}{2\pi }\,.
\end{equation}
Having the eigenvalue density, we can compute the expectation value of the circular Wilson loop and the free energy. We find:
\be\label{weakWloop}
W(C) = 1+\frac{\lambda _R}{8}+\frac{\lambda _R^2}{192}+O\left(\lambda _R^3\right)
= 1+\frac{1}{8}\,\lambda+\left( \frac{1}{192}+\frac{\KK'(M)}{64\pi^2}\right) \lambda^2 +O(\lambda^3).
\ee
For the derivatives of the free energy, given by (\ref{dF}), we have:
\bea
\label{freeJ}
\frac{\partial F}{\partial\lambda} &=& -\frac{\lambda _R}{2\lambda }
\\
 \frac{\partial F}{\partial M}&=& -\mathcal{K}(M)-\frac{\lambda _R}{16\pi ^2}\,\mathcal{K}''(M),
\eea
where in the first equation we used
\be
\left\langle x^2\right\rangle = \frac{\mu^2}{4}=\frac{\lambda _R}{16\pi ^2}\,,
\ee
and in the second equation expanded the kernel to the second order in $x-y$. Integrating we obtain:
\be
F = -\frac{1}{2}\,\ln \lambda_R +\ln H(M)+O\left(\lambda _R^2\right)
= -\frac{1}{2}\,\ln \lambda +\ln H(M)
-\frac{\KK'(M)}{16\pi^2}\, \lambda +O\left(\lambda ^2\right).
\label{weakF}
\ee
In particular, using the asymptotic behavior of $H(x)$ (see appendix~\ref{kernelappendix}), we note that at large $MR$ (here we recover the dependence on $R$): 
$$
 \ln Z \to N^2 M^2R^2 \,\ln MR \qquad   (MR\gg 1).
$$
This reproduces the expected UV divergence of the partition function originating from zero modes of the one-loop determinant\footnote{We thank Arkady Tseytlin
for comments on this point.}.

One can notice from the calculations above that the perturbative expansion is reorganized in power series in $\lambda _R$. At $M\sim 1$ this is not a big change, as $\lambda _R\sim \lambda $, but when $M$ becomes big, $\lambda _R$ contains a large logarithm. When written in terms of $\lambda _R$, expressions above resum large logs of the form $\lambda (\lambda \ln M)^n$ and, for the Wilson loop, $\lambda ^2(\lambda \ln M)^n$.  The limit of $M\rightarrow \infty $ at fixed $\lambda _R$ corresponds to $\mathcal{N}=2$ SYM with fixed dynamical scale $\Lambda $. The results above are valid when the radius of the sphere is small in the $\mathcal{N}=2$ units, and consequently $\lambda _R\ll 1$. At finite radius (finite $\lambda _R$), the width of the eigenvalue distribution $\mu $ is no longer small compared to one, but is still much smaller than $M$. The integral equation valid in this regime is obtained from (\ref{nnstar}) by expanding only the last two terms in the kernel:
\begin{equation}
 \label{nn2}
\strokedint_{-\mu}^\mu dy \rho(y) 
\left(\frac{1}{x-y} -\KK(x-y)\right)= \frac{8\pi^2}{\lambda_R}\ x.
\end{equation}
This equation describes pure $\mathcal{N}=2$ SYM on $S^4$ and was analyzed in detail in \cite{Russo:2012ay}\footnote{\label{importantfootnote}In comparing to \cite{Russo:2012ay} it is necessary to take into account that our definition of $\Lambda $ differs from that in \cite{Russo:2012ay} by a factor of $\,{\rm e}\,^{-1-\gamma }$.}.


\subsection{Strong coupling}

The limit of $\lambda \gg 1$ is supposed to have a holographic description in terms of a weakly-curved supergravity dual. For the theory defined on flat $\mathbbm{R}^4$ the dual geometry is known \cite{Pilch:2000ue}. The large-$N$ solution of $\mathcal{N}=2^*$ SYM at strong coupling \cite{Buchel:2013id}, continued to infinite volume of $S^4$, completely agrees with available predictions of holography. We briefly review these results, for completeness.

As $\lambda $ is increased, the attractive linear force $8\pi^2x/\lambda $ in the saddle-point equation (\ref{nnstar}) becomes weaker, and the eigenvalue distribution expands to larger and larger values of $x$. Eventually, at sufficiently big $\lambda $, the width of the eigenvalue distribution becomes much larger than the bare mass scale of the $\mathcal{N}=2^*$ theory: $\mu \gg M$. In this case we can assume that $|x-y|\gg M$ and rescale $x\rightarrow x\mu $. This justifies the approximation\footnote{It is also true that $|x-y|\gg 1$, which is important for the last step in the approximation.}
\begin{equation}
 \frac{1}{2}\,\KK(x-y+M)+\frac{1}{2}\,\KK(x-y-M)-\KK(x-y)\approx 
 \frac{1}{2}\,\KK''(x-y)M^2\approx \frac{M^2}{x-y}\,.
\end{equation}
At the end, the only effect of the complicated $\mathcal{K}$-terms in the equations is a multiplicative renormalization of the Hilbert kernel:
\begin{equation}
 \frac{1}{x-y}~\longrightarrow~ \frac{1+M^2}{x-y}\ .
\end{equation}
The equation  is again solved by the Wigner distribution (\ref{WignerLaw}), but now with
\begin{equation}
\label{strongmu}
 \mu =\frac{\sqrt{\lambda \left(1+M^2\right)}}{2\pi}\,.
\end{equation}

The Wilson loop vev then behaves as 
\begin{equation}\label{W(C)atlargelambda}
 \ln W(C)\simeq  \sqrt{\lambda \left(1+M^2\right)},
\end{equation}
and the free energy is given by\footnote{We shifted the result in \cite{Buchel:2013id} by a constant in order to uniformly normalize the free energy across the whole phase diagram, and in particular to make it agree with the free energy of $\mathcal{N}=4$ SYM at $M=0$, in the normalization of \cite{Russo:2012ay}. Of course adding a constant to the free energy does not change any physics.}
\begin{equation}\label{Fatlargel}
 F=-\frac{1+M^2}{2}\,\ln\frac{\lambda \left(1+M^2\right)\,{\rm e}\,^{2\gamma +\frac{1}{2}}}{16\pi ^2}+\gamma +\frac{1}{4}-\ln 4\pi .
\end{equation}

The strongly-coupled $\mathcal{N}=2^*$ theory, on flat space, is dual to type IIB supergravity on the Pilch-Warner background \cite{Pilch:2000ue}. In the equations above, we can reach the flat-space limit by rescaling $M\rightarrow MR$, $\mu \rightarrow \mu R$ and taking $R\rightarrow \infty $, which amounts to just dropping the $1$ in $1+M^2$. The results are then in perfect agreement with supergravity. In particular, the probe analysis of the Pilch-Warner solution shows that the eigenvalues of the Higgs vev are distributed on an interval in the two-dimensional moduli space with the semicircular Wigner density \cite{Buchel:2000cn}. The width of the distribution, $\mu =\sqrt{\lambda }M/2\pi $, is the same as (\ref{strongmu}) with $M\gg 1$. The semicircular shape of the eigenvalue distribution is actually a generic prediction of the probe analysis, valid for many other supergravity backgrounds \cite{Carlisle:2003nd}. It is a challenge for localization to reproduce this universality on the field-theory side. 

The vev of the circular Wilson loop (\ref{W(C)atlargelambda}) obeys the perimeter law at strong coupling. Extrapolating this behavior to generic Wilson loops, we may expect that the Wilson loop vev for a  contour of length $L$ behaves as
$\ln W(C)\simeq \sqrt{\lambda }LM/2\pi $, assuming $LM\gg 1$. Taking the standard value $\sqrt{\lambda }/2\pi $ for the dimensionless string tension, and using the same regularization prescription as in $AdS_5\times S^5$, one reproduces this result from the minimal area law in the Pilch-Warner geometry \cite{Buchel:2013id}.
To compare the supergravity predictions with localization for finite $MR$, it is necessary to find the supergravity solution whose boundary is $S^4$ rather than flat $\mathbbm{R}^4$. First steps towards constructing such solution were taken in \cite{Buchel:2003qm,Buchel:2013fpa}. 

\subsection{Conformal perturbation theory}

In the limit of mass going to zero, the theory flows to ${\cal N}=4$ SYM. The matrix model then becomes Gaussian, and the eigenvalues form the Wigner distribution (\ref{WignerLaw}) of width $\mu =\sqrt{\lambda }/2\pi $. If $\lambda $ is large, the circular Wilson loop \cite{Erickson:2000af,Drukker:2000rr} and the free energy \cite{Russo:2012ay} computed in the Gaussian matrix model agree with the minimal area law and the on-shell action of type IIB supergravity on $AdS_5\times S^5$. It is of interest to calculate the first correction in $M^2$, at any $\lambda $. The expansion in $M$ can be interpreted as conformal perturbation theory.

Expanding (\ref{nnstar}) to the leading order in $M$, we get:
\begin{equation}
 \strokedint_{-\mu }^{\mu }dy\,\,\frac{\rho (y)}{x-y}
 =\frac{8\pi ^2}{\lambda }\,x-\frac{M^2}{2}\int_{-\mu }^{\mu }
 dy\,\rho (y)\mathcal{K}''(x-y).
\end{equation}
Treating the second term on the right hand side as perturbation we can plug in the leading order solution for the $\rho (y)$ and, using the Fourier representation (\ref{FourierKprpr}),  we obtain
\begin{equation}
 \strokedint_{-\mu }^{\mu }dy\,\,\frac{\rho (y)}{x-y}
 =\frac{8\pi ^2}{\lambda }\,x-\frac{4\pi M^2}{\sqrt{\lambda }}\int_{0 }^{\infty  }
 d\omega \,\,\frac{\omega J_1\left(\frac{\sqrt{\lambda }\omega }{\pi }\right)\sin 2\omega x }{\sinh^2\omega }\,.
\end{equation}
The Hilbert kernel is inverted by applying
 \be
 \strokedint_{-\mu}^ \mu  \frac{dx}{\sqrt{\mu^2-x^2}} \frac{1}{z-x}
 \ee
 to both sides of the equation. We thus find:
\be\label{rhosmallM}
\rho(x) = \frac{8\pi }{\lambda }\,\sqrt{\mu^2-x^2}\left[ 1 -\frac{\sqrt{\lambda }M^2}{\pi }
\int_{0 }^{\infty  }
 d\omega \,\int_{0}^{1}ds\,\,\frac{\omega^2 J_1\left(\frac{\sqrt{\lambda }\omega }{\pi }\right)J_0\left(\frac{\sqrt{\lambda }\left(1-s\right)\omega }{\pi }\right)
 \cos 2s\omega x }{\sinh^2\omega }
 +O\left(M^4\right)
\right].
\ee

To determine $\mu $, we need to impose the normalization condition on the density. Integrating both sides of  (\ref{rhosmallM}) from $-\mu $ to $\mu $, and replacing $\mu $ by $\sqrt{\lambda }/2\pi $ in the second term we get:
\begin{equation}
 1=\frac{4\pi ^2\mu ^2}{\lambda }
 -2M^2
 \int_{0 }^{\infty  }
 d\omega \,\int_{0}^{1}ds\,\,\frac{\omega J_1\left(\frac{\sqrt{\lambda }\omega }{\pi }\right)J_0\left(\frac{\sqrt{\lambda }\left(1-s\right)\omega }{\pi }\right)
 J_1\left(\frac{\sqrt{\lambda }s\omega }{\pi }\right)
  }{s\sinh^2\omega }
 +O\left(M^4\right).
\end{equation}
The $s$-integral here can be computed explicitly, and we finally obtain:
\begin{equation}\label{intrepformu}
 \mu =\frac{\sqrt{\lambda }}{2\pi }\left(
 1+M^2\int_0^\infty d\omega \,\,\frac{\omega J^2_1\left(\frac{\sqrt{\lambda }\omega }{\pi }\right)}{\sinh^2\omega }
 +O\left(M^4\right)
 \right)
\end{equation}

This expression is first order in $M^2$, but is non-perturbative in $\lambda $. To make contact with the results of the two previous sections, we consider the limiting cases of $\lambda \ll 1 $ and $\lambda \gg 1$.  If $\lambda $ is large, the main contribution to the integral comes from $\omega \sim 1/\sqrt{\lambda }$. The $\sinh^2\omega $ in the denominator can be then approximated by $\omega ^2$, after which the whole expression integrates to $1/2$.
 We thus find
\be
\mu \simeq \frac{\sqrt{\lambda}}{2\pi }\ \left(1+\frac{M^2}{2}\right)\qquad \left(\lambda \rightarrow \infty \right),
\ee
in agreement with the strong-coupling result (\ref{strongmu}) expanded to the first order in $M^2$.

The weak-coupling limit is even simpler, we just need to expand the Bessel function under the integral. This gives 
\be
\mu \simeq \frac{\sqrt{\lambda}}{2\pi}\ \left(1+\frac{3\,\zeta(3)\lambda M^2}{8\pi^2}\right)\qquad \left(\lambda \rightarrow 0\right),
\ee
which matches with (\ref{mmaa}), (\ref{renormalizedl}), if we take into account that $\mathcal{K}'(M)=6\,\zeta (3)M^2+O(M^4)$ according to (\ref{zeta-expansion}).

The weak-coupling expansion in fact can be carried out to all orders in $\lambda $:
\begin{equation}\label{mu=series}
 \mu =\frac{\sqrt{\lambda }}{2\pi }\left[1+M^2
 \sum_{k=1}^{\infty }\left(-1\right)^{k+1}\frac{\left(2k\right)!\left(2k+1\right)!\,\zeta \left(2k+1\right)}{\left(k-1\right)!\left(k+1\right)!k!^2}\left(\frac{\lambda }{16\pi ^2}\right)^k
 +O\left(M^4\right)\right].
\end{equation}
It is necessary to stress that the general arguments on the structure of perturbation series given in the introduction are not valid in the conformal limit. These arguments should apply in the opposite, IR regime, when the sphere is big and effective field theory gives an accurate description of physics. The $
M^2$ correction, calculated above, should be interpreted as the first order of conformal perturbation theory around the $\mathcal{N}=4$ point. As such, it should inherit the perturbative structure of the $\mathcal{N}=4$ SYM, well understood due to integrability \cite{Beisert:2010jr}. 

In a finite theory, such as $\mathcal{N}=4$ SYM, planar perturbation theory should have a finite radius of convergence \cite{Brezin:1977sv} which is determined by the combinatorics of planar graphs and thus should not depend much on the particular observable. Since the spectrum of local gauge-invariant operators in $\mathcal{N}=4$ SYM is quite well understood, we can draw some conclusions on the radius of convergence from the spectral problem. Computation of the anomalous dimensions of local operators can be conveniently mapped to a spin-chain problem \cite{Minahan:2002ve,Beisert:2003tq,Beisert:2003yb}. Due to integrability, the only necessary input is the dispersion relation of the elementary magnon excitations of the spin chain and their two-body S-matrix. The exact dispersion relation of the $\mathcal{N}=4$ magnon is \cite{Beisert:2004hm,Beisert:2005tm}
\begin{equation}
 \epsilon _{\rm mag}(p)=\sqrt{1+\frac{\lambda }{\pi ^2}\,\sin^2\frac{p}{2}}.
\end{equation}
This expression has manifestly finite radius of convergence in $\lambda $. The  ``staggered" magnon with momentum at the edge of the Brillouin zone: $p=\pi $ has the smallest radius of convergence. Its energy has a square root branch point at 
\begin{equation}\label{lambdac}
 \lambda _c=-\pi ^2.
\end{equation}
 This should be the radius of convergence of generic observable.
 
 Quite remarkably, the radius of convergence of perturbation series in (\ref{mu=series}) is exactly the same. Indeed, perturbative coefficients  in (\ref{mu=series}) behave as $\,{\rm const}\,\cdot \pi ^{-2k}$ at large $k$ indicating a pole at $\lambda =-\pi ^2$. From the point of view of the integral representation (\ref{intrepformu}), the singularity at $\lambda =-\pi ^2$ occurs because of the exponential growth of the Bessel function of an imaginary argument, which saturates the convergence of the integral at large $\omega $ when $\lambda $ approaches $\lambda _c$. Using the large-argument asymptotics of the Bessel function we get:
\begin{equation}
 \mu \simeq \frac{\sqrt{\lambda }M^2}{\pi ^2+\lambda }\qquad \left(\lambda \rightarrow -\pi ^2\right).
\end{equation}

The quantities that may have a more direct $\mathcal{N}=4$ interpretation are the free energy and the circular Wilson loop, to the  computation of which we now proceed.
To calculate the free energy we integrate (\ref{rhosmallM}) with the $x^2$ weight, which gives:
\begin{equation}
 \left\langle x^2\right\rangle=\frac{\pi ^2\mu ^4}{\lambda }+\frac{M^2}{2}
 \int_{0 }^{\infty  }
 d\omega \,\int_{0}^{1}ds\,\,\frac{ J_1\left(\frac{\sqrt{\lambda }\omega }{\pi }\right)J_0\left(\frac{\sqrt{\lambda }\left(1-s\right)\omega }{\pi }\right)
  }{\omega \sinh^2\omega }
  \,\,
  \frac{\partial^2 }{\partial s^2}
  \,\,
  \frac{J_1\left(\frac{\sqrt{\lambda }s\omega }{\pi }\right)}{s}
 +O\left(M^4\right).
\end{equation}
Taking $\mu $ from (\ref{intrepformu}), calculating the $s$-integral, and using (\ref{dF}), we get:
\begin{equation}\label{dFdlambda}
 \frac{\partial F}{\partial \lambda }=-\frac{1}{2\lambda }
 -\frac{8\pi ^2M^2}{\lambda^2 }
 \int_{0 }^{\infty  }
 d\omega \,\,\frac{  J_1\left(\frac{\sqrt{\lambda }\omega }{\pi }\right)
  \left( J_1\left(\frac{\sqrt{\lambda }\omega }{\pi }\right) -\frac{\sqrt{\lambda }\omega }{2\pi }\, J_0\left(\frac{\sqrt{\lambda }\omega }{\pi }\right)\right)}{\omega  \sinh^2\omega }
   +O\left(M^4\right).
\end{equation}
This equation can be integrated to
\begin{equation}\label{xxactF}
 F=-\frac{1}{2}\,\ln\lambda -\frac{4\pi ^2M^2}{\lambda }
  \int_{0 }^{\infty  }
  d\omega \,\,\frac{ \frac{\lambda \omega^2 }{4\pi ^2}-J^2_1\left(\frac{\sqrt{\lambda }\omega }{\pi }\right)}{ \omega \sinh^2\omega }
   +O\left(M^4\right).
\end{equation}
The leading term here is the free energy of $\mathcal{N}=4$ SYM. It can be understood holographically \cite{Russo:2012ay} as arising from the log-divergence of the on-shell supergravity action on $AdS_5\times S^5$ upon taking into account  a factor of $\sqrt{\lambda }$  in the radius-energy relation  \cite{Peet:1998wn,Bianchi:2001de} in the AdS/CFT correspondence.

Again, we can check consistency with the weak-coupling results by expanding the Bessel function:
\begin{equation}
 F\simeq -\frac{1}{2}\ln\lambda -\frac{3\zeta \left(3\right)\lambda M^2}{8\pi ^2}\qquad \left(\lambda \rightarrow 0\right).
\end{equation}
This is in agreement with
(\ref{weakF}), when (\ref{zeta-expansion}) is used for $\ln H(M)$ and $\mathcal{K}'(M)$. Checking the strong-coupling limit is a bit trickier, since the integral in (\ref{xxactF}) logarithmically diverges on the upper limit if $\sinh^2\omega $ is replaced by $\omega ^2$. 
Cutting off the resulting integral at $\omega \sim  1$ we get, with the logarithmic accuracy:
\begin{equation}
 F\simeq -\frac{1}{2}\left(1+M^2\right)\ln\lambda,
\end{equation}
in precise agreement with (\ref{Fatlargel}).  Systematic strong-coupling expansion of (\ref{xxactF}) requires matching contributions from $\omega \sim 1/\sqrt{\lambda }$ and $\omega \sim 1$. To the first two non-vanishing orders,
\begin{equation}
 \left.\frac{\partial F}{\partial M^2}\right|_{M=0}
 =-\frac{1}{2}\,\ln\frac{\lambda }{16\pi ^2}-\frac{3}{4}-\gamma -\frac{2\pi ^2}{3\lambda }+O\left(\frac{1}{\lambda ^{\frac{3}{2}}}\right).
\end{equation}
In general the expansion goes in powers of $1/\sqrt{\lambda }$, as expected from string theory in $AdS_5\times S^5$. It is curious though that the term of order $M^2/\sqrt{\lambda }$ is absent.

The radius of convergence of perturbation series for the free energy is also $\pi ^2$ as in  (\ref{lambdac}). The singularity at $\lambda = -\pi ^2$ is a logarithmic branch cut:
\begin{equation}
 F= {\rm analytic}-\frac{8M^2}{\pi ^3}\left(\pi ^2+\lambda \right)\ln\left(\pi ^2+\lambda \right)\qquad \left(\lambda \rightarrow -\pi ^2\right).
\end{equation}

Finally, one can also compute the vev of the circular Wilson loop, $W=\langle e^{2\pi x}\rangle $. It is useful to consider a more general expectation value:
\begin{eqnarray}
 \left\langle \,{\rm e}\,^{2\pi ax}\right\rangle
 &=&\frac{4\pi \mu I_1\left(2\pi a\mu \right)}{a\lambda }-M^2\int_{0}^{\infty }
 d\omega \,\,\frac{\omega ^2J_1\left(\frac{\sqrt{\lambda }\omega }{\pi }\right)}{\sinh^2\omega }
 \int_{0}^{1}ds\,J_0\left(\frac{\sqrt{\lambda }\left(1-s\right) \omega}{\pi }\right)\left[
 \frac{J_1\left(\frac{\sqrt{\lambda }s\omega }{\pi }+ia\sqrt{\lambda }\right)}{s\omega +i\pi a}
 \right.
\nonumber \\
&&\left.
 +\left(a\rightarrow -a\right)
 \vphantom{\frac{J_1\left(\frac{\sqrt{\lambda }s\omega }{\pi }+\frac{ia\sqrt{\lambda }}{2\pi }\right)}{2s\omega +ia}
}
 \right],
\end{eqnarray}
which can be brought to a simpler form:
\begin{equation}
 \left\langle \,{\rm e}\,^{2\pi ax}\right\rangle
 =\frac{2I_1\left(a\sqrt{\lambda }\right)}{a\sqrt{\lambda }}+2\pi aM^2
 \int_{0}^{\infty }
 d\omega \,\,\frac{\omega J_1\left(\frac{\sqrt{\lambda }\omega }{\pi }\right)
 \left(
 \pi aI_0\left(a\sqrt{\lambda }\right) J_1\left(\frac{\sqrt{\lambda }\omega }{\pi }\right)
-\omega  I_1\left(a\sqrt{\lambda }\right) J_0\left(\frac{\sqrt{\lambda }\omega }{\pi }\right)
 \right)}{\left(\omega ^2+\pi ^2a^2\right)\sinh^2\omega }\,.
\end{equation}
Eq.~(\ref{dFdlambda}) can be obtained by expanding this expression to  second order in $a$. Setting $a=1$, we get:
\begin{equation}\label{confptforW}
 W(C)
 =\frac{2I_1\left(\sqrt{\lambda }\right)}{\sqrt{\lambda }}+2\pi M^2
 \int_{0}^{\infty }
 d\omega \,\,\frac{\omega J_1\left(\frac{\sqrt{\lambda }\omega }{\pi }\right)
 \left(
 \pi I_0\left(\sqrt{\lambda }\right) J_1\left(\frac{\sqrt{\lambda }\omega }{\pi }\right)
-\omega  I_1\left(\sqrt{\lambda }\right) J_0\left(\frac{\sqrt{\lambda }\omega }{\pi }\right)
 \right)}{\left(\omega ^2+\pi ^2\right)\sinh^2\omega }\,.
\end{equation}
It is straightforward to check that the weak-coupling result (\ref{weakWloop}) is reproduced when $W(C)$ is expanded in $\lambda $. The leading asymptotics at strong coupling is
\begin{equation}
 W(C)\simeq \sqrt{\frac{2}{\pi }}\,\lambda ^{-\frac{3}{4}}\,{\rm e}\,^{\sqrt{\lambda }}\left(1+\frac{\sqrt{\lambda }M^2}{2}\right)\qquad \left(\lambda \rightarrow \infty \right),
\end{equation}
again in agreement with the previous result, eq.~(\ref{W(C)atlargelambda}). Near the critical point $\lambda =-\pi ^2$, the Wilson loop, like the free energy, has a logarithmic branch point:
\begin{equation}
 W(C)\simeq -4M^2J_1\left(\pi \right)\ln\left(\pi ^2+\lambda \right)
 \qquad 
 \left(\lambda \rightarrow -\pi ^2\right).
\end{equation}

The circular Wilson loop in $\mathcal{N}=4$ theory is given by the first term in (\ref{confptforW}) \cite{Erickson:2000af,Drukker:2000rr} and has an infinite radius of convergence, which happens because of massive diagram cancellations. It is known that only rainbow graphs without internal vertices contribute \cite{Erickson:2000af}.
In this sense the circular  Wilson loop at $M=0$ is not a generic observable.
At the first order of conformal perturbation theory, the circular loop starts to receive contributions from generic diagrams. This shifts the radius of convergence to the expected $\lambda _c=-\pi^2$ point. 
Making a more direct contact with the AdS/CFT integrability, and in particular reformulating conformal perturbation theory for the free energy (\ref{xxactF}) and for the Wilson loop (\ref{confptforW}) in the $\mathcal{N}=4$ language is an interesting open problem.

\subsection{Decompactification limit} \label{decompact:sec}

The compactification on the sphere can be considered just as a convenient way to impose an IR cutoff, necessary for computing the path integral by localization. If we take this point of view, the radius of the sphere $R$ should be sent to infinity at the end of the calculation. All the dimensionful parameters, including $\mu $ and $M$, scale linearly with $R$ and after $R$ is sent to infinity and eliminated from the equations, the dimensionful quantities regain their canonical dimensions. In the limit $R\rightarrow \infty $, the Hilbert kernel drops out from the saddle-point equation (\ref{nnstar}), and  $\mathcal{K}(x)$ can be approximated by its asymptotics at infinity (\ref{KKapprox}).  It is convenient to differentiate the resulting equation twice, which gives:
\begin{equation}\label{intnormal1}
 \strokedint_{-\mu }^\mu
 dy\,\rho (y)  \left( \frac{2}{x-y} - \frac{1}{x-y+M}-\frac{1}{x-y-M} \right) =0.
\end{equation}
The Hilbert kernel of the original equation produces an $1/(x-y)^3$ term, which scales as $1/R^2$ and can therefore be neglected in the decompactification limit. 

As shown in \cite{Russo:2013qaa}, the boundary conditions at the ends of the interval   change from the square root at finite $R$ to the inverse square root in the strict $R\rightarrow \infty $ limit. The integral equation (\ref{intnormal1}) with the inverse-square-root boundary conditions has normalizable solutions for any $\mu $ (the norm can be adjusted by simply multiplying $\rho (x)$ with a constant). An extra condition, which fixes $\mu $, follows from the integrated form of (\ref{intnormal1}), equivalent to the original saddle-point equation differentiated once:
\begin{equation}\label{formintegrated}
 \int_{-\mu }^{\mu }dx\,\rho (x)\ln\left(\frac{M^2}{x^2}-1\right)^2=\frac{16\pi ^2}{\lambda }\,.
\end{equation}

The set of equations (\ref{intnormal1}), (\ref{formintegrated}) can be easily solved at weak coupling, when $\mu \ll M$. The last two terms in the kernel then approximately cancel leading to a very simple equation
\begin{equation}
  \strokedint_{-\mu }^\mu
 dy\,\rho (y) \,\frac{1}{x-y}=0,
\end{equation}
whose properly normalized solution is
\begin{equation}\label{simpleinversesqrt}
 \rho (x)=\frac{1}{\pi \sqrt{\mu ^2-x^2}}\qquad \left(\lambda \rightarrow 0\right).
\end{equation}
An extra condition (\ref{formintegrated}) then determines $\mu $:
\begin{equation}\label{muatverysmalll}
 \mu =2M\,\,{\rm e}\,^{-\frac{4\pi ^2}{\lambda  }}=2\Lambda \qquad \left(\lambda \rightarrow 0\right),
\end{equation}
where $\Lambda $ is the dynamically generated scale in the IR limit of the pure $\mathcal{N}=2$ theory. The solution (\ref{simpleinversesqrt}) was derived from localization in \cite{Russo:2012ay} and reproduces earlier results obtained by taking the large-$N$ limit within Seiberg-Witten theory \cite{Douglas:1995nw,Ferrari:2001mg}. Interestingly, the same distribution of eigenvalues arises in one of the supergravity solutions proposed as a holographic dual of pure $\mathcal{N}=2$ SYM \cite{Bigazzi:2001aj}.

The system of equations (\ref{intnormal1}), (\ref{formintegrated}) can be solved analytically without making any approximations \cite{Russo:2013qaa} with the help of the method proposed in \cite{Hoppe,Kazakov:1998ji}. The solution is actually valid as long as $\lambda $ is not too big. When $\lambda $ reaches a critical value  $\lambda_c\approx 35$, the theory undergoes a transition to a new phase. Similar phase transitions happen at larger couplings: $\lambda _c^{(2)}\approx 83$, $\lambda _c^{(3)}\approx 150$, and so on, with an infinite sequence of critical points accumulating at strong coupling. 

The phase transitions are caused by new light states that appear in the spectrum. Each pole in the kernel of the integral equation (\ref{intnormal1}) corresponds to a massless, or nearly massless particle. The pole at $x=y$ arises due to the photons of the unbroken $U(1)^{N-1}$, while the poles at $x=y\pm M$ correspond to massless hypermultiplets\footnote{More precisely, to very light hypermultiplets, whose masses scale as $1/N$ and $1/N^2$ in the large-$N$ limit (cf.~\cite{Douglas:1995nw}).}. Of course at weak coupling, when $M\gg \mu $, all hypermultiplets are quite heavy. This is reflected in the integral equation (\ref{intnormal1}) by the absence of hypermultiplet poles in the region of integration, where  $|x-y|<M$ for any $x$ and $y$ lying within the interval $(-\mu ,\mu )$. The largest possible value of $|x-y|$ is equal to $2\mu $.  When $\lambda $ becomes bigger, $\mu $ also grows and eventually exceeds $M/2$. When $\mu$ reaches $M/2$ the first resonance appears in the spectrum, causing transition to a new phase. At $\mu =M$ the secondary resonance appears, leading to another phase transition, and so on. The $n$-th critical point is determined by the condition
\begin{equation}\label{mucn}
 \mu \left(\lambda _c^{(n)}\right)=\frac{nM}{2}\,.
\end{equation}
Using the strong-coupling solution (\ref{strongmu}) we can estimate  asymptotically
\begin{equation}
 \lambda _c^{(n)}\simeq \pi ^2n^2\qquad \left(n\rightarrow \infty \right).
\end{equation}

After reviewing the exact solution found in \cite{Russo:2013qaa} in the weak-coupling phase, we will study the critical behavior near the first phase transition, and then will analyze the structure of the strong-coupling phase in more detail.

\subsubsection{Exact solution in weak-coupling phase}

The solution found in \cite{Russo:2013qaa} is written down in terms of the resolvent:
\begin{equation}\label{resolvent}
 G(z)=\int_{-\mu }^{\mu }\frac{dy\,\rho (y)}{\left(z-y\right)^2-\frac{M^2}{4}}\,.
\end{equation}
The resolvent is an analytic function on the complex plane with two distinct cuts $(\pm M/2-\mu ,\pm M/2+\mu)$, centered at $\pm M/2$. When $\lambda $ is small, the cuts are very short and are well separated. 
With $\lambda $ growing, the endpoints of the  cuts move closer to the origin and eventually collide at $z=0$, after which the two cuts coalesce. That happens when $\mu =M/2$.
This is another way to see how the phase transition arises in the solution of the saddle-point equations. 

The eigenvalue density can be found from the resolvent by taking discontinuity across one of the cuts:
\begin{equation}\label{densityasdisc}
 \rho (x)=\frac{M}{2\pi i}\left(G\left(x+\frac{M}{2}-i0\right)-G\left(x+\frac{M}{2}+i0\right)\right).
\end{equation}
Qualitatively, it has the same shape as (\ref{simpleinversesqrt}), with the inverse square root singularities at the endpoints and a minimum at $z=0$, although the precise functional form of the density is more complicated than a simple square root.
 
As long as the cuts do not overlap, that is, before the phase transition  the resolvent can be found exactly by integrating the equation \cite{Russo:2013qaa}:
\begin{equation}\label{dz-dG}
 dz=-\frac{dG}{2\sqrt{G^3\left(1+\xi G\right)\left(1+\eta G\right)\left(1+\bar{\eta }G\right)}}\,.
\end{equation}
The inverse function $z(G)$ can thus be expressed in terms of elliptic integrals.
The parameters that characterize the solution\footnote{Those are related to the parameters $\xi _i$  in \cite{Russo:2013qaa} as follows: $\xi_1=-\xi  $, $\xi _2=-\bar{\eta} $, $\xi _3=\eta $.} are given by 
\begin{eqnarray}\label{xi-eta-throughtheta}
 \xi &=&\frac{M^2}{12}\left(E_2+\theta _3^4+\theta _4^4\right)
\nonumber \\
\eta &=&\frac{M^2}{12}\left(E_2-2\theta _3^4+\theta _4^4\right)
\nonumber \\
\bar{\eta }&=&\frac{M^2}{12}\left(E_2+\theta _3^4-2\theta _4^4\right),
\end{eqnarray}
where $E_2\equiv E_2(-r^2)$ is the Eisenstein series of index two, and $\theta _a\equiv \theta _a(0|ir)$ are the theta-constants, which both depend on the 't~Hooft coupling through the modular parameter
\begin{equation}\label{modularr}
 r=\,{\rm e}\,^{-\frac{4\pi ^2}{\lambda }}.
\end{equation}
Notations and conventions for the theta-functions and Eisenstein series are listed in the appendix~\ref{thetappendix}, where we also collect some of their useful properties. An equivalent representation in terms of elliptic integrals is given in appendix~\ref{appendixcriticalcalculations}. It is clear that the resolvent only depends on symmetric combinations of the three parameters $\xi $, $\eta $, $\bar{\eta }$. As a consequence, correlation functions $<x^{2n}>$ are symmetric polynomials in $\xi $, $\eta $, $\bar{\eta }$ of degree $n$ \cite{Russo:2013qaa}. It is shown in appendix~\ref{thetappendix} that any such symmetric polynomial can be expressed  through the Eisenstein series only, with all theta-constants canceling out.

The width of the eigenvalue distribution is given by
\begin{equation}\label{muastheta}
\mu = -\frac{iM}{2} \frac{\theta_4'(v)}{\theta_4(v)}\,,
\ee
where the parameter $v$ is a solution of the transcendental equation
\begin{equation}\label{transeqforv}
 \frac{\theta _1^2(v)}{\theta _4^2(v)}=\frac{2\theta _3^4-\theta _4^4-E_2}{3\theta _2^2\theta _3^2}\,.
\end{equation}

The theta-functions of a real argument and pure imaginary modulus satisfy 
$$\overline{\theta ^4_{1,2}}=-\theta ^4_{1,2},\qquad \overline{\theta ^4_{3,4}}=\theta ^4_{4,3}.
$$
Consequently, $\eta $ and $\bar{\eta }$ are complex conjugate to one another and $\xi $ is real positive. It is more difficult to see that $\mu $ is real and positive. 
One can show using Landen transformations that the solution of (\ref{transeqforv}) is of the form $v=\pi /4+is$ with real negative $s$ and that for such $v$, $\mu $ given by (\ref{muastheta}) is indeed real  \cite{Russo:2013qaa}.

Expanding (\ref{muastheta}), (\ref{transeqforv}) at small $\lambda $ we get:
\begin{equation}
 \frac{\mu}{M}  = 2 \,{\rm e}\,^{-\frac{4 \pi ^2}{\lambda }}-4 \,{\rm e}\,^{-\frac{12 \pi ^2}{\lambda }}
-16 \,{\rm e}\,^{-\frac{28 \pi   ^2}{\lambda }} -58 \,{\rm e}\,^{-\frac{36 \pi ^2}{\lambda }}-324 \,{\rm e}\,^{-\frac{44 \pi ^2}{\lambda}}
-1856 \,{\rm e}\,^{-\frac{52 \pi ^2}{\lambda }}+\ldots 
\label{+glambdaexpanded}
\end{equation}
The first term reproduces (\ref{muatverysmalll}). The rest of the expansion has precisely the form (\ref{weak-expansion}) anticipated on general grounds from the OPE in the effective field theory. The OPE coefficients can in principle be computed to any desired order and are just numbers (potentially they could have been power series in $\lambda $). 

For the free energy, the OPE can actually be resummed into a rather compact expression. 
The free energy can be calculated from (\ref{dF}), using
\begin{equation}\label{<x^2>}
 \left\langle x^2\right\rangle=\frac{\xi +\eta +\bar{\eta }}{3}-\frac{M^2}{12}
 =\frac{M^2}{12}\left(E_2-1\right).
\end{equation}
The first equality follows from comparing the Laurent expansion of the resolvent (\ref{resolvent}) at $z=\infty $ with that of the exact solution (\ref{dz-dG}). The second equality is a consequence of (\ref{xi-eta-throughtheta}). It is straightforward to generalize this computation to higher correlators $\langle x^{2n}\rangle$ with $n>1$. Explicit expressions for the first few are listed in appendix~\ref{thetappendix}. Integrating  (\ref{dF}) in $\lambda $, we find:
\begin{equation}
\label{freefe}
F= -M^2\ln \left(M\,{\rm e}\,^{\gamma -\frac{1}{2}}\prod_{n=1}^{\infty }\left[1-\left(-1\right)^n\,{\rm e}\,^{-\frac{8\pi ^2n}{\lambda }}\right]^2\right).
\ee
Up to an unessential, $\lambda $-independent constant the free energy can be written as
\begin{equation}
 F=\,{\rm const}\,+2M^2\sum_{k=1}^\infty  (-1)^{k}\sigma_{-1}(k)  \,{\rm e}\,^{-\frac {8\pi^2 k}{\lambda}}.
\end{equation}
The divisor function $\sigma_{-1}(k)$ has a nice combinatorial interpretation, which suggests that the quantum field theory calculation reproducing the OPE coefficients for the free energy
might reduce to simple combinatorics. 

The above formulas are non-perturbative in $\lambda $, and have remarkably simple modular properties. It is actually tempting to extend them to strong coupling with the help of modular transformations (see appendix~\ref{thetappendix}). Unfortunately, this  makes little sense, as the strong-coupling regime is described by a totally different solution. The phase transition that happens in between invalidates analytic continuation to $\lambda $ larger than $\lambda_c $. We now turn to the detailed discussion of the phase transition.

\subsubsection{Phase transition}

The phase transition, according to (\ref{mucn}), happens when $\mu (\lambda _c)=M/2$.
As shown in \cite{Russo:2013qaa}, the parameter $\xi $ vanishes at the critical point\footnote{This happens because $G(0)=-1/\xi $. At the critical point the two cuts of the resolvent collide at $z=0$, and since the resolvent is singular at the endpoints of the cuts, $G(0)$ blows up.}: $\xi (\lambda _c)=0$. The critical coupling thus corresponds to the solution of the transcendental equation $E_2+\theta _3^4+\theta _4^4=0$. Solving this equation numerically we find:
\begin{equation}\label{rclc}
  r_c=0.328107\ldots \qquad \lambda _c=35.4252\ldots  
\end{equation}

A convenient small parameter in the vicinity of the critical point is
\begin{equation}\label{Deltadef}
 \Delta =\mu -\frac{M}{2}\,.
\end{equation}
In appendix~\ref{appendixcriticalcalculations} we show that in the weak-coupling phase
\begin{equation}\label{Delta-xi}
 \Delta =-\frac{2\xi ^{\frac{3}{2}}}{3|\eta |}+o\left(\xi ^{\frac{3}{2}}\right),
\end{equation}
or, in terms of the coupling constant:
\begin{equation}\label{Deltathroughlambda}
 \Delta \simeq -C_\Delta M\left(\lambda _c-\lambda \right)^{\frac{3}{2}}\ ,
\end{equation}
with
\begin{equation}\label{CDelta}
 C_\Delta =\frac{2\sqrt{2}\pi ^3\theta _3^4\theta _4^4}{3\lambda _c^3}
 =0.0031347\ldots
\end{equation}
The theta-constants here are evaluated at the critical coupling given by (\ref{rclc}).  

The critical density $\rho _c(x)$ (the eigenvalue density right at the critical point) can be calculated by setting $\xi =0$ in (\ref{dz-dG}). Even then the resolvent is still  an elliptic integral. However, certain simplifications occur near the endpoints of the eigenvalue distribution,  when the resolvent becomes very large and the $1$'s  in $(1+\eta G)$ and $(1+\bar{\eta }G)$ can be dropped. Then (\ref{dz-dG}) can be easily integrated:
\begin{equation}
 G(z)\simeq \left(3|\eta |\right)^{-\frac{2}{3}}\left(z-M\right)^{-\frac{2}{3}}.
\end{equation}
This approximation is valid as long as $|z-M|\ll M$. The constant of integration was chosen such that the branch point lies exactly at $z=\mu_c +M/2=M$. Taking discontinuity of the resolvent across the cut, we find for the density, according to (\ref{densityasdisc}):
\begin{equation}\label{critical density}
 \rho _c(x)\simeq {C_\rho }{M^{-\frac{1}{3}}\left(\frac{M}{2}-x\right)^{-\frac{2}{3}}}\ ,
\end{equation}
with
\begin{equation}\label{constantCrho}
 C_\rho =\frac{M^{\frac{4}{3}}}{2\cdot 3^\frac{1}{6}\pi |\eta |^{\frac{2}{3}}}
 =\frac{2^{\frac{1}{3}}}{3^{\frac{1}{6}}\pi \theta _3^{\frac{4}{3}}\theta _4^{\frac{4}{3}}}=0.198421\ldots 
\end{equation}
Interestingly, and in contradistinction to more conventional matrix models, the endpoint singularity hardens at the critical point. The scaling exponent changes from $-1/2$ away from criticality to $-2/3$ at the transition point.

\subsubsection{Critical behavior}\label{critical section}

It was found numerically \cite{Russo:2013qaa} that after the phase transition the density develops two cusps at  $x=\pm (M- \mu ) $.
The dynamical reason  can be understood from the saddle-point equation (\ref{intnormal1}): 
when $\mu>M/2$ the force due to the $1/(x-y \pm M)$ terms in the equation becomes repulsive on the interval between the poles at $x=\pm (M- \mu ) $ and the edge of the eigenvalue distribution. This force pushes  eigenvalues toward $x=\pm (M- \mu ) $ forming the cusps. 

We did not succeed in finding an analytic solution for the eigenvalue density in the strong-coupling phase, but as a first step we can study the critical behavior in the vicinity of the transition point, where the equations  substantially simplify.

Away from the critical point, but still in the vicinity of the phase transition, the density should not be too different from $\rho_c$. In the bulk of the eigenvalue distribution, the difference is linear in $\lambda_c-\lambda \sim |\Delta| ^{2/3} $. The deviation must be parametrically larger near the endpoints, to accommodate the change in the endpoint exponents. 
The characteristic scale in the vicinity of the endpoints is $|\Delta |$, 
making
\begin{equation}
 u=\frac{\frac{M}{2}-x}{|\Delta |}
\end{equation}
the appropriate scaling variable.
We will study the density in the limit of $\Delta \rightarrow 0$ with $u$ finite. It is reasonable to expect that the density assumes a universal shape in this scaling regime.

 We introduce two scaling functions, $f_+(u)$ and $f_-(u)$, which describe the endpoint behavior of the density respectively above and below the phase transition, such that near  $x=\mu $ the density assumes the following form:
\begin{equation}\label{scdensity}
 \rho (x)=C_\rho M^{-\frac{1}{3}}|\Delta |^{-\frac{2}{3}}f_\pm\left(\frac{\frac{M}{2}-x}{|\Delta |}\right),\qquad \pm=\mathop{\mathrm{sign}}\Delta .
\end{equation}
The constant $C_\rho $ is defined in (\ref{constantCrho}) and is introduced here to match with the critical solution (\ref{critical density}).
The scaling functions $f_\pm(u)$ are defined on the semi-infinite intervals $(\mp 1,\infty )$, and should not depend on any parameters. Away from the endpoints the density asymptotes to the critical solution. Comparing (\ref{scdensity}) with (\ref{critical density}), we find that at large $u$:
\begin{equation}\label{finfinity}
 f_\pm(u)\simeq u^{-\frac{2}{3}}\qquad \left(u\rightarrow \infty \right).
\end{equation}
Both functions have an inverse square-root singularity at the endpoints of the intervals on which they are defined: $f_{\pm}\sim 1/\sqrt{u\pm 1}$. The function $f_+$, in addition, is expected to have a cusp at $u=1$.

We first compute the scaling function $f_-(u)$, which can be extracted from the exact solution (\ref{dz-dG}) in the weak-coupling phase. To this end, we define the scaling resolvent:
\begin{equation}
 g(w)=\frac{\sqrt{3}}{2^{\frac{5}{3}}\pi }\int_{1}^{\infty }\frac{dv\,f_-(v)}{v-w}\,.
\end{equation}
The overall numerical factor is introduced for future convenience. The resolvent has a cut from $1$ to $\infty $ along the real axis, with discontinuity equal to the scaling function:
\begin{equation}\label{discont of g}
 g(u\pm i0)=-\frac{r(u)}{2}\pm 2^{-\frac{5}{3}}\sqrt{3}\,if _-(u) \qquad \left(u\in\mathbbm{R},\,u>1\right).
\end{equation}

On the other hand, we can substitute the scaling form of the density (\ref{scdensity}) into the definition of the full resolvent (\ref{resolvent}) to find the relationship between the latter and its scaling form:
\begin{equation}\label{GGgg}
 G\left(M-|\Delta |w\right)=\left(\frac{2}{3|\eta |\Delta }\right)^{\frac{2}{3}}g(w)=\frac{1}{\xi }\,g(w)\ ,
\end{equation}
where we assume that $w\sim 1$ at $\Delta \rightarrow 0$, and used (\ref{constantCrho})  and  (\ref{Delta-xi}) to express the coefficient of proportionality in terms of $\xi $, one of the constants that defines the exact solution (\ref{dz-dG}).

In the scaling limit the constant $\xi $ goes to zero, while two other constants, $\eta $ and $\bar{\eta }$, stay finite. We see that the product $\xi G$ also remains finite, which means  that $\eta G,\bar{\eta }G\gg 1$. In this approximation, (\ref{dz-dG}) can be explicitly integrated:
\begin{equation}
 3|\eta ||\Delta |w=\left(2\xi G-1\right)\sqrt{\frac{\xi G+1}{G^3}}\,.
\end{equation}
Using (\ref{GGgg}) and  once again (\ref{Delta-xi}), we arrive at the cubic equation for $g(w)$:
\begin{equation}\label{cubic}
 4\left(w^2-1\right)g^3+3g=1.
\end{equation}
This equation is universal, it does not depend on any parameters.
Setting $w=u\pm i0$, and substituting (\ref{discont of g}) in (\ref{cubic}) we find
\begin{eqnarray}\label{solf-}
 &&f_-(u)=2^{\frac{2}{3}}\sqrt{\frac{1}{u^2-1}+r^2}
\nonumber 
 \\
 &&4\left(u^2-1\right)r^3+3r=1.
\end{eqnarray}

It is easy to see that the scaling function has the correct end-point behavior:
\begin{equation}
 f_-(u)\simeq \frac{2^{\frac{1}{6}}}{\sqrt{u-1}}\qquad \left(u\rightarrow 1^+\right).
\end{equation}
As expected, $f_-(u)$ asymptotes to the critical solution $u^{-2/3}$ at infinity:
\begin{equation}
 f_-(u)\simeq u^{-\frac{2}{3}}\qquad \left(u\rightarrow \infty \right).
\end{equation}

Interestingly, a closed equation for $f_\pm(u)$ can be obtained by taking the scaling limit of (\ref{intnormal1}). This is possible because the density has an inverse square root singularity at the endpoints and, as soon as $x$ is close enough to $\mu $, the contribution from the bulk of the eigenvalue distribution scales away in the $\Delta \rightarrow 0$ limit. Substituting the scaling form of the density (\ref{scdensity}) into the integral equation, we obtain:
\begin{equation}\label{eqforf}
 \strokedint_{\mp 1}^\infty dv\,f_\pm(v)\left(\frac{2}{v-u}+\frac{1}{v+u}\right)=0,
\end{equation}
which has to be solved with the boundary condition (\ref{finfinity}).

The resulting equation is reminiscent of the saddle-point equations in the exactly solvable $O(n)$ matrix model \cite{Kostov:1988fy,Gaudin:1989vx,Kostov:1992pn,Eynard:1992cn,Eynard:1995nv,Eynard:1995zv,Chekhov:1996xy} with $n=1$.  The  $O(1)$ model reduces to solving a cubic equation \cite{Eynard:1992cn}, quite similar to the formula (\ref{solf-}) for $f_-$ that we obtained by indirect arguments. Neither these arguments, nor, seemingly, a more direct approach of  \cite{Eynard:1992cn,Eynard:1995nv} can be  used to solve for $f_+$, because of the cusp singularity in the middle of the eigenvalue distribution. It is nevertheless possible to understand the analytic structure of the cusp without finding an exact solution.

The integral equation for $f_+(u)$ can be simplified by a substitution\footnote{Suggested to us by D.~Volin.}
\begin{equation}\label{Volin's surmise}
 f_+(u)=
\begin{cases}
 \frac{4}{3}\,h(u)+\frac{2}{3}\,h(-u) & {\rm }1>u>-1
\\
 h(u)  & {\rm } u>1.
\end{cases}
\end{equation}
If the function $h(u)$ satisfies
\begin{equation}\label{hequation}
 2\strokedint_{-1}^\infty dv\,\,\frac{h(v)}{u-v}=\int_{1}^{\infty }dv\,
 \frac{h(v)}{u+v}\,,
\end{equation}
the function $f_+(u)$, constructed via (\ref{Volin's surmise}), can be checked to satisfy the integral equation (\ref{eqforf}).  The boundary condition on $f_+(u)$  translates to the asymptotic behavior $h(u)\simeq u^{-2/3}$.

The right-hand side of the integral equation (\ref{hequation}) is a continuous function on the whole interval $(-1,\infty )$. Its inverse Hilbert transform is consequently also continuous. Therefore,  $h(u)$ has no singularities anywhere between $-1$ and $\infty $. This becomes clear once the Hilbert kernel in (\ref{hequation}) is inverted:
\begin{equation}
 h(u)=\frac{1}{2\pi \sqrt{u+1}}\int_{1}^{\infty }dv\,\,\frac{\sqrt{v-1}\,h(v)}{u+v}\,.
\end{equation}
The correct boundary conditions at $u=-1$ and $u=\infty $ also become manifest. Indeed, $h(u)\sim 1/\sqrt{u+1}$ at $u\rightarrow -1$. Assuming that $h(u)\simeq \,{\rm const}\, \cdot  u^{-\alpha }$ at $u\rightarrow \infty $, we find that $2\sin\pi (\alpha -1/2)=1$ and thus $\alpha =2/3$. The unique solution with $\,{\rm const}\,=1$ can be obtained by iterations, which actually converge rather fast. Solving the equation numerically, we find that $h(u)$ is a monotonously decreasing function and $h(u)\simeq 3C_+(u+1)^{-1/2}/2$  at $u\rightarrow 1$ with
\begin{equation}
 C_+=0.650367\ldots  
\end{equation}

Although $h(u)$ is continuous on the interval $(-1,\infty )$, the change of variables (\ref{Volin's surmise}) induces a cusp singularity in $f_+(u)$ at $u=1$. The structure of the cusp is easy to infer from (\ref{Volin's surmise}): to the left of the cusp, $f_+$ has an inverse square root singularity, while from the right  it approaches a finite limiting value given by
\begin{equation}
 \tilde{C}_+=h(1)=0.531021\ldots 
\end{equation}
The scaling function thus has the following singularity structure:
\begin{equation}\label{singularities of f+}
 f_+(u)\simeq 
\begin{cases}
 \frac{2C_+}{\sqrt{u+1}} & {\rm } (u\rightarrow -1^+)
 \\
\frac{C_+}{\sqrt{1-u}} 
   & {\rm }(u\rightarrow 1^-)
   \\
  \tilde{C}_+ & {\rm }(u\rightarrow 1^+).
   \end{cases}
\end{equation}
\begin{figure}[t]
\begin{center}
 \centerline{\includegraphics[width=11cm]{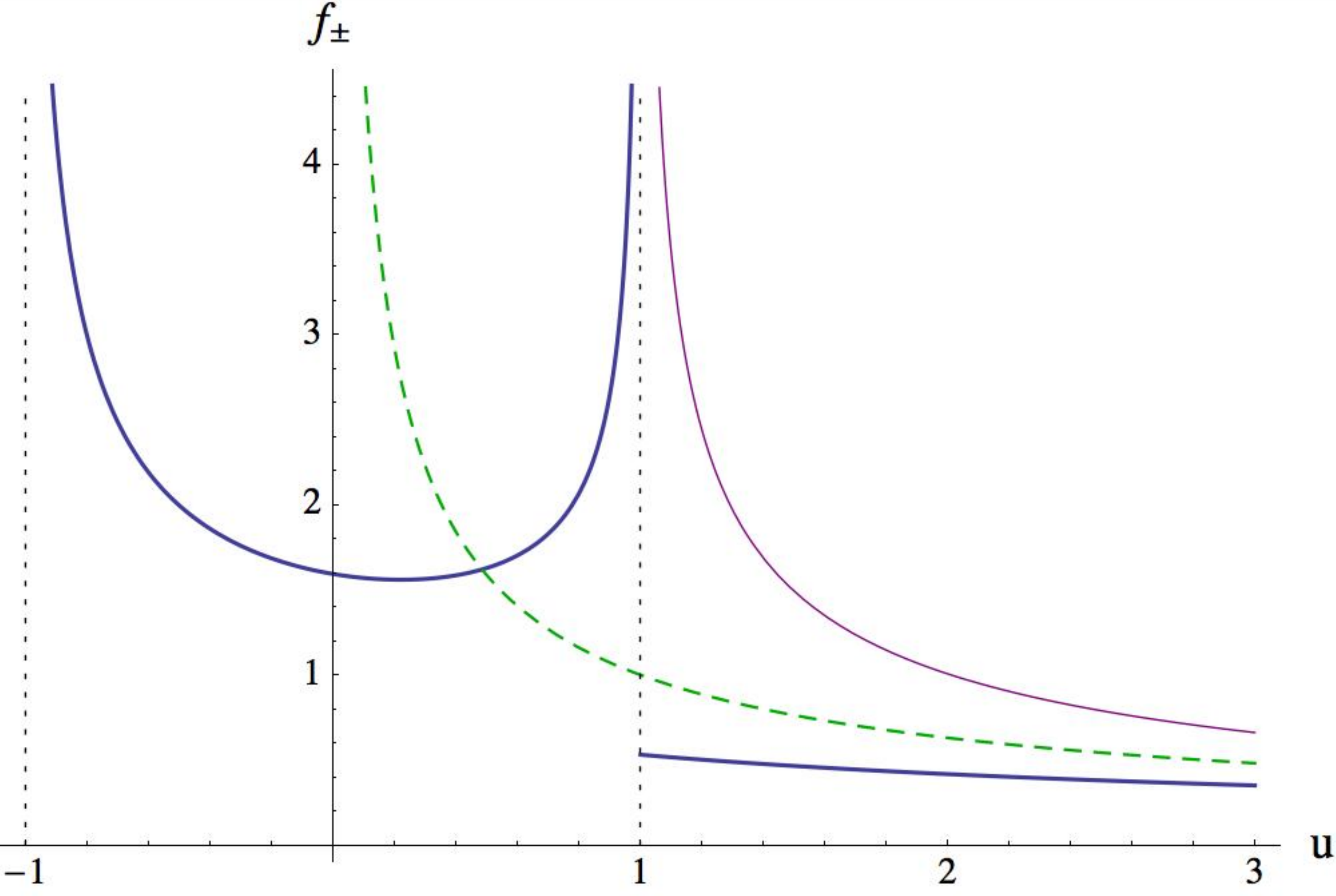}}
\caption{\label{pictureoffs}\small The scaling functions $f_+(u)$ (thick blue line) and $f_-(u)$ (thin purple line). Also shown is the critical solution $u^{-2/3}$ (dashed green line).}
\end{center}
\end{figure}
The scaling functions $f_+$ and $f_-$ are depicted in fig.~\ref{pictureoffs}.

\begin{figure}[t]
\begin{center}
 \centerline{\includegraphics[width=11cm]{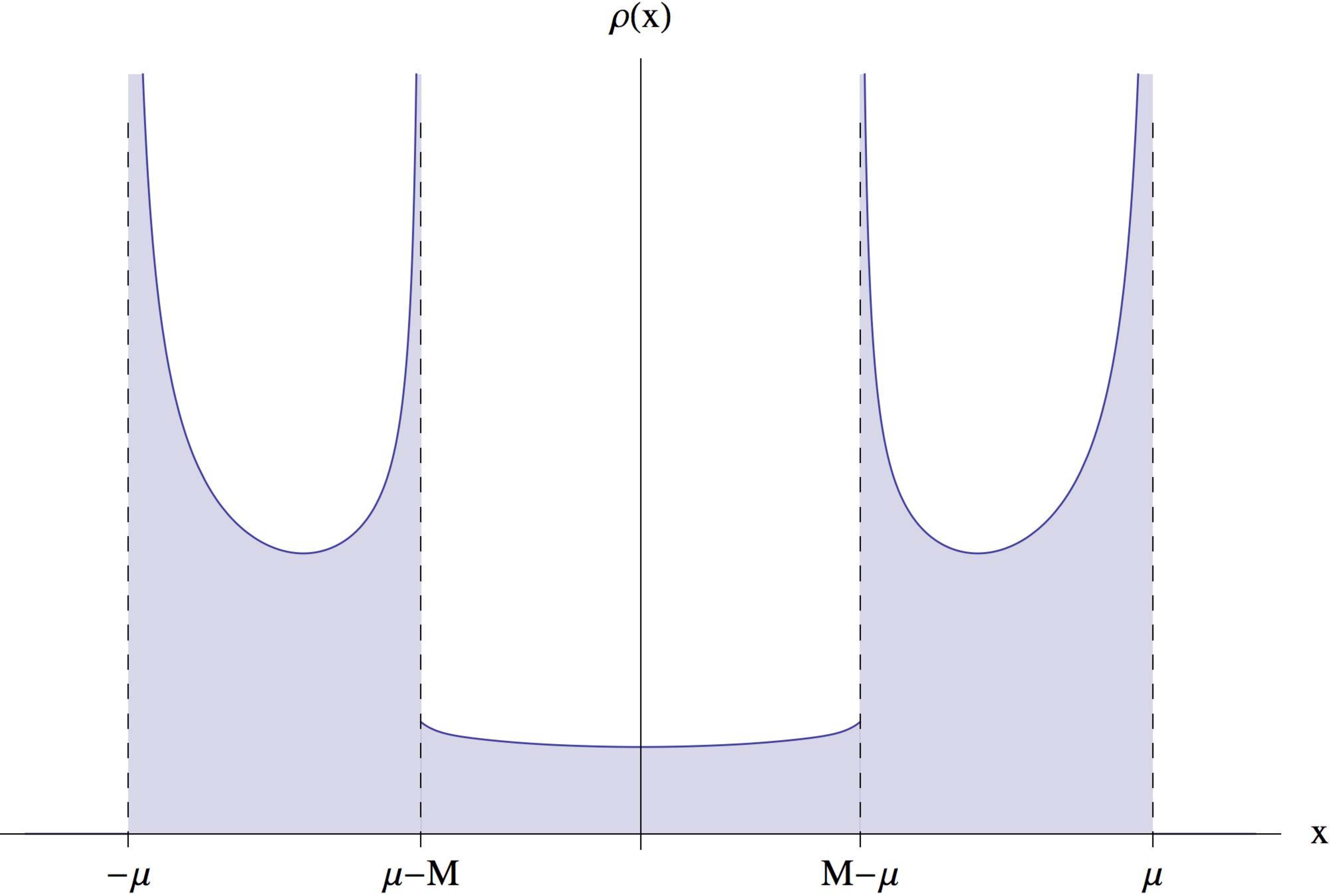}}
\caption{\label{afterphase}\small The eigenvalue density in the strong-coupling phase. The plot shows the numerical solution of the saddle-point equations at  $\mu =0.7M$, $\lambda \approx 54$.}
\end{center}
\end{figure}
Finite distance away from the phase transition we can solve the saddle-point equations numerically. The structure of the density remains qualitatively the same all the way up to the next critical point. The density has two cusps at $\pm (M-\mu )$ of the following structure: the density is finite on the inside of the cusp, and diverges as an inverse square root on the outside (fig.~\ref{afterphase}).  The cusp is in resonance with the opposite edge of the eigenvalue density (the distance corresponds exactly to the massless hypermultiplet). The repulsive force, that only acts on the eigenvalues between the cusp and the other edge of the eigenvalue density, causes the pileup on the outside of the cusp. Since the cusp is an image of the edge, the pileup is in general weaker than the edge singularity as can be seen from eq.~(\ref{singularities of f+}), or from fig.~\ref{afterphase}.

\subsubsection{Remarks on thermodynamics.}

An interesting question to address is the order of the phase transition. The thermodynamic singularity at the transition point  can be conveniently characterized by the second moment of the eigenvalue density $\left\langle x^2\right\rangle$, 
which according to (\ref{dF}) plays the r\^ole of
a heat capacity. In the weak-coupling phase $\left\langle x^2\right\rangle$ is given by eq.~(\ref{<x^2>}). 
\begin{figure}[t]
\begin{center}
 \centerline{\includegraphics[width=10cm]{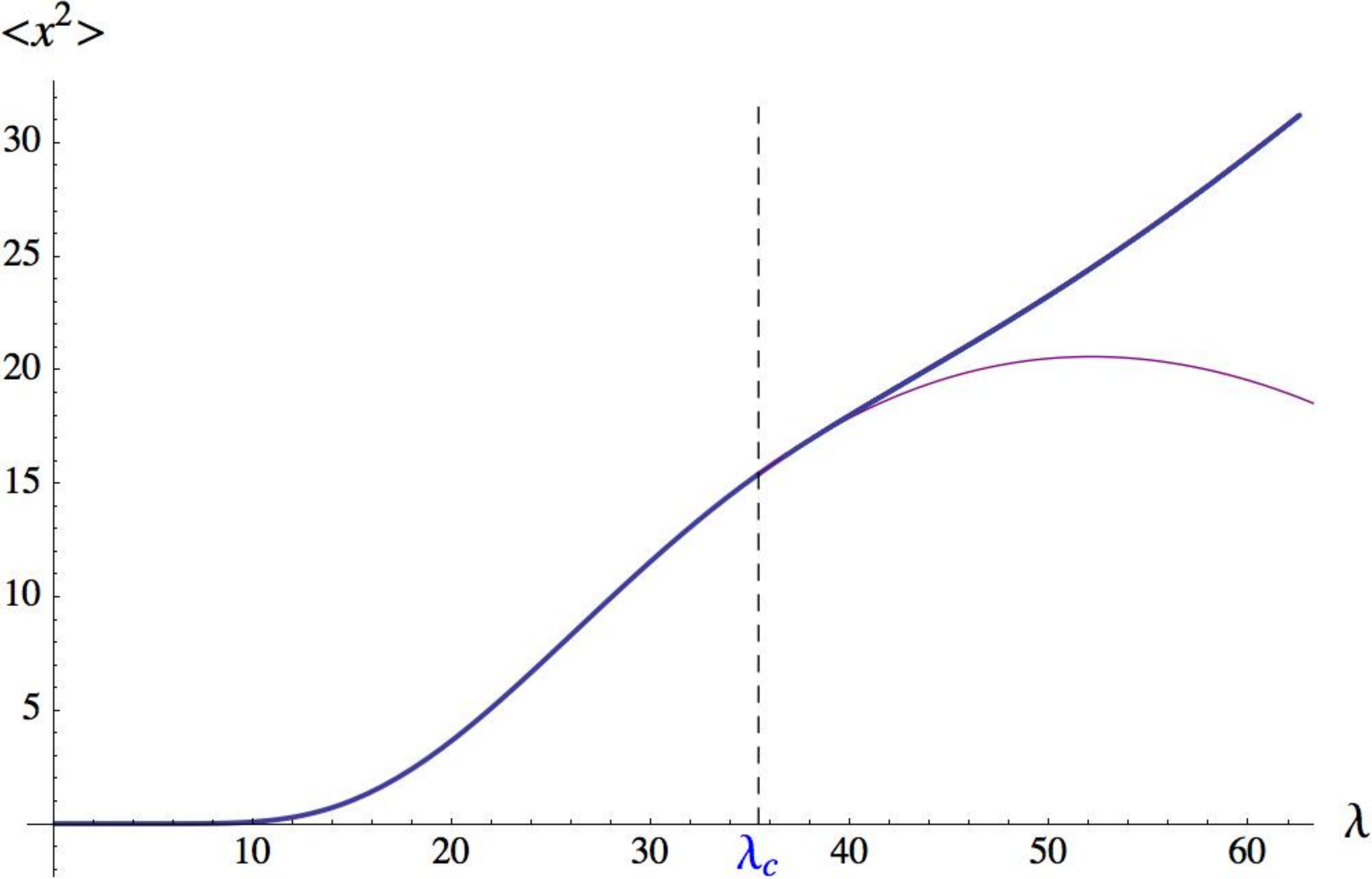}}
\caption{\label{figx2}\small The second moment of the eigenvalue density as a function of the 't~Hooft coupling for $M=10$. Below the phase transition the free energy is calculated analytically from (\ref{<x^2>}). The curve above the transition  is obtained numerically. The analytic continuation of the weak-coupling result past $\lambda _c$ is shown as the thin purple line.}
\end{center}
\end{figure}
Above the transition we computed it numerically. We plot these results in fig.~\ref{figx2}.
It is clear from the plot that the transition is rather smooth. 
\begin{figure}[t]
\begin{center}
 \centerline{\includegraphics[width=10cm]{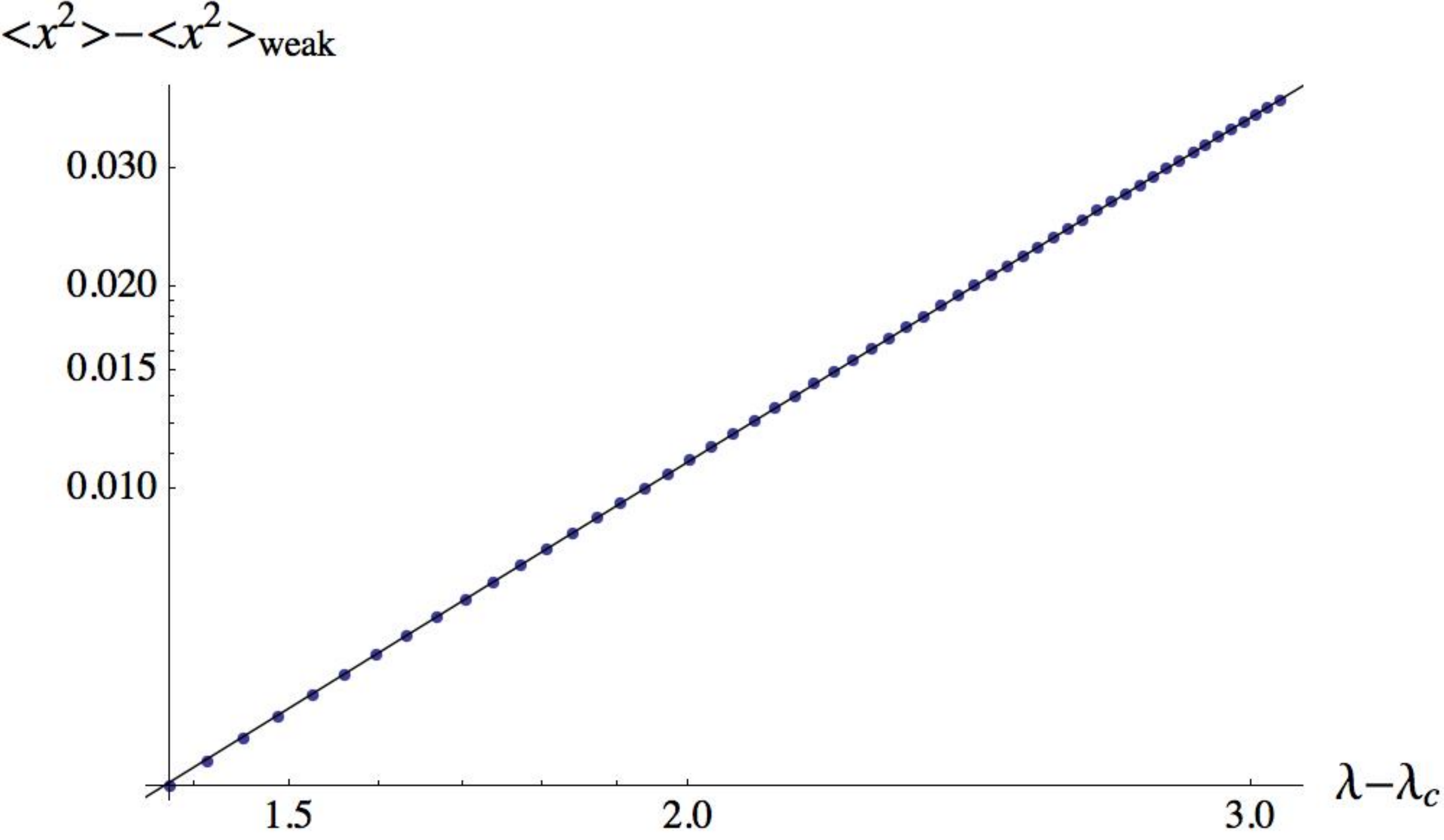}}
\caption{\label{log-log}\small The log-log plot of the deviation of the exact $\left\langle x^2\right\rangle$ from the analytic continuation of the weak-coupling result. The dots are obtained by numerically solving the saddle-point equation. The solid line is a fit to $a(\lambda -\lambda _c)^3+b(\lambda -\lambda _c)^4$. The best-fit values of the coefficients are $a\simeq 1.5\cdot 10^{-3}$, $b\simeq 5\cdot 10^{-5}$. $M$ is set to $10$.}
\end{center}
\end{figure}

The log-log plot of the difference between the true $\left\langle x^2\right\rangle$  and $\left\langle x^2\right\rangle_{\rm weak}$, the analytic continuation of the weak-coupling result past the transition point, is shown in fig.~\ref{log-log}. As expected the difference grows as a power\footnote{Our accuracy is insufficient to exclude possible logarithmic corrections to the power-law scaling.} of the distance to the critical point:  $\left\langle x^2\right\rangle-\left\langle x^2\right\rangle_{\rm weak}\sim \left(\lambda -\lambda _c\right)^\alpha $. Within the numerical precision of our calculations, the scaling exponent  is consistent with $\alpha =3$, which  indicates that the first two derivatives of $f'(\lambda )\sim \left\langle x^2\right\rangle$ are continuous, while the third derivative experiences a finite jump:
\begin{equation}
 \left\langle x^2\right\rangle-\left\langle x^2\right\rangle_{\rm weak}
 \simeq C_2M^2\left(\lambda -\lambda _c\right)^3\qquad \left(\lambda \rightarrow \lambda _c^+\right).
\end{equation}
For the constant $C_2$ we get an estimate $C_2\simeq 1.5\cdot 10^{-5}$. We thus conclude that the transition is of the fourth order with zero critical exponents. The fourth derivative of the free energy experiences a finite jump at the transition point.

In this regard it is interesting to look at higher derivatives of the free energy in the weak-coupling phase. They can all  be computed analytically. 
For convenience we consider $\beta\equiv 8\pi^2/\lambda $ as an independent variable.
The first derivatives of the free energy in $\beta $ follows from (\ref{dF}), (\ref{<x^2>}):
\begin{equation}
 \frac{\partial F}{\partial \beta} = \frac{M^2}{12} \left( E_2 -1\right),
\end{equation}
where the argument of $E_2$ is $-\,{\rm e}\,^{-\beta }$. Using the standard  Ramanujan identities for differentiation of Eisenstein series, we further find:
\bea
\frac{\partial ^2F}{\partial \beta^2} &=&   \frac{M^2}{144} \left( E_4- E_2^2  \right)
\nonumber\\
 \frac{\partial ^3F}{\partial \beta^3}& =&  \frac{M^2}{864}\left( E_2^3 - 3E_2E_4 +2E_6 \right)
\eea
It then follows from (\ref{utiles}) that 
\be
M^4 \ \frac{d^3F}{d\beta^3} =  2 \xi\eta\bar\eta 
\ee
The parameter $\xi $ vanishes at the critical point and, with it, the third derivative of the free energy. The critical point can thus be identified with the inflection point 
of $E_2(-\,{\rm e}\,^{-\beta })$.

Another interesting quantity is the order parameter of the phase transition. Although no real order parameter can exist, as no symmetry is broken at the transition point,   the transition is nevertheless caused by  IR physics, and we can try to identify a parameter that controls the correlation length that diverges. The first guess is  $1/\xi $, since $\xi $ vanishes at the critical point. Let us recall the definition of $\xi $:
\begin{equation}
 \frac{1}{\xi }=G(0)=\left\langle \frac{1}{x^2-\frac{M^2}{4}}\right\rangle.
\end{equation}
The hypermultiplet masses are proportional to  $|x-y \pm  M|$, and in this sense, $\xi $ is an average mass squared of the hypermultiplet, with averaging defined in a specific way. At the transition point, the first massless hypermultiplet appears in the spectrum and $1/\xi $ diverges. Consequently,  $1/\sqrt{\xi }$ represents the correlation length and $\sqrt{\xi }$ can be indeed identified with the mass gap that closes at the point of phase transition.
\begin{figure}[t]
\begin{center}
 \centerline{\includegraphics[width=10cm]{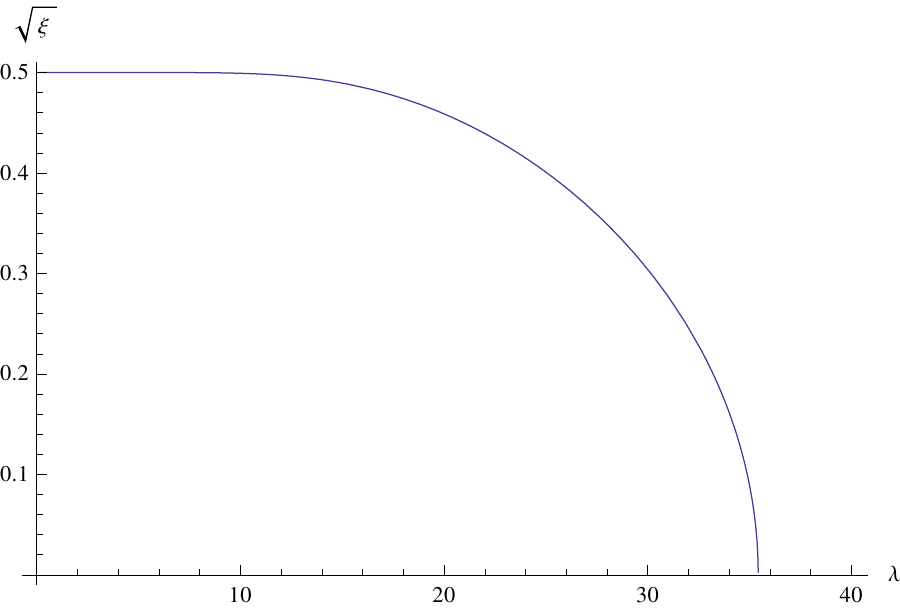}}
\caption{\label{sqrt-xi}\small $\sqrt{\xi}$ as a function of  $\lambda $  at $M =10$.}
\end{center}
\end{figure}
Fig.~\ref{sqrt-xi} shows $\sqrt{\xi}$ as a function of  $\lambda $. As can be seen from (\ref{Delta-xi}), (\ref{Deltathroughlambda}), $\sqrt{\xi }$ vanishes at the critical point with the critical exponent $1/2$.

As the coupling $\lambda $ is increased, 
the system undergoes an infinite sequence of  phase transitions occurring whenever $\mu $ crosses  thresholds at $nM/2$, with integer $n$.
All transitions are of a similar nature, with a pair of new cusps created at the boundary of the eigenvalue distribution. The different phases are described  in more detail in \cite{Russo:2013qaa}. As the coupling grows the cusps proliferate, but become less and less pronounced. The enveloping curve of the limiting density at very strong coupling approaches the Wigner semicircular shape, reproducing the result (\ref{WignerLaw}), (\ref{strongmu}) of the strong-coupling  analysis, albeit in a somewhat irregular fashion, since on top of the enveloping curve the density has a complicated non-analytic fine structure.

\subsection{Arbitrary $\lambda $ and $M$}

We have so far analyzed the saddle-point equations (\ref{nnstar}) 
near the edges of the phase diagram in fig.~\ref{phaseportrait}. Although solving for the density in the analytic form is a difficult problem, the solution can be constructed numerically with good accuracy for any given $M$ and $\lambda $. Let us describe qualitatively the behavior of the eigenvalue density across the whole phase diagram.

\begin{figure}[t]
\begin{center}
 \centerline{\includegraphics[width=8cm]{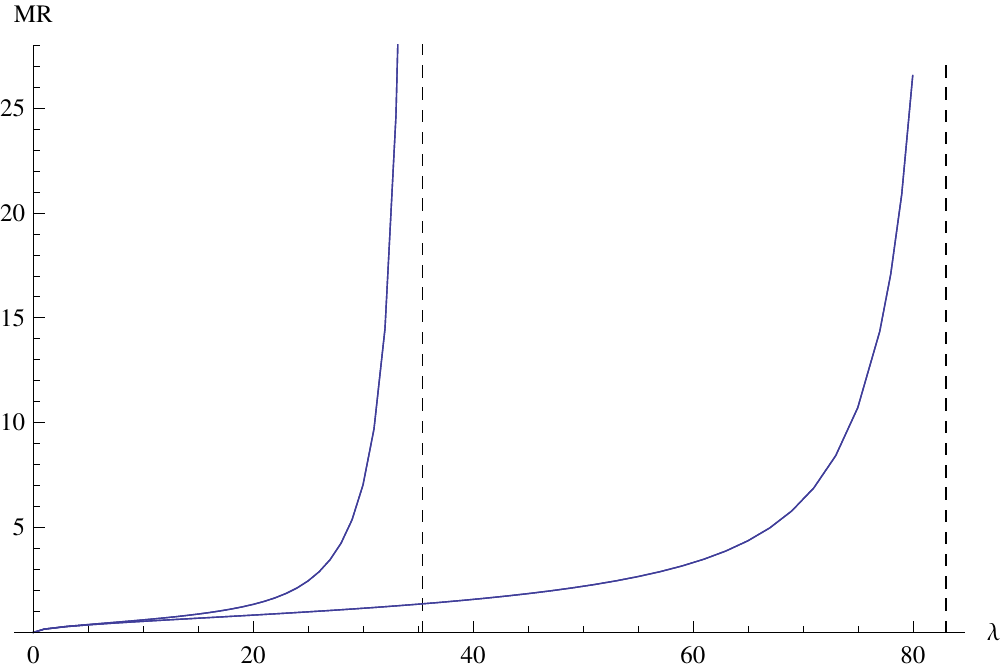}}
\caption{\label{sch2}\small The lines $\mu=M/2$ and $\mu=M$. The former approaches $M=\infty $ at $\lambda \approx 35.4$, the latter approaches $M=\infty $ at $\lambda\approx 83$, in consistency with the phase transitions found at these values of $\lambda $ for the decompactified ($MR=\infty $) theory. }
\end{center}
\end{figure}
The eigenvalues tend to spread more and more with  $M$ or $\lambda $ growing: $\partial \mu /\partial M>0$ and $\partial \mu /\partial \lambda >0$. The  $\mu =M/2$ and $\mu =M$ contours in the $\lambda -M$ plane are displayed in fig.~\ref{sch2}. These contours can be regarded as  crossover lines, since at $M\rightarrow \infty $ they approach the critical values of the coupling $\lambda =\lambda _c^{(1)}$, $\lambda =\lambda _c^{(2)}$, at which the infinite-volume system undergoes phase transitions. The phase transitions disappear at finite $M$,
since the IR effects responsible for the critical behavior are regulated by the finite volume of the four-sphere  (recall that $M$ should be understood as  $MR$). At very large but finite $M$,  $\mu =Mn/2$ are the crossovers lines, on which the fourth derivative of the free energy rapidly changes. Likewise, the cusps in the eigenvalue density become sharp peaks of finite height. These features  quickly disappear away from the infinite-volume limit and are not really visible at moderate values of $M$.

\begin{figure}[t]
\begin{center}
 \centerline{\includegraphics[width=11cm]{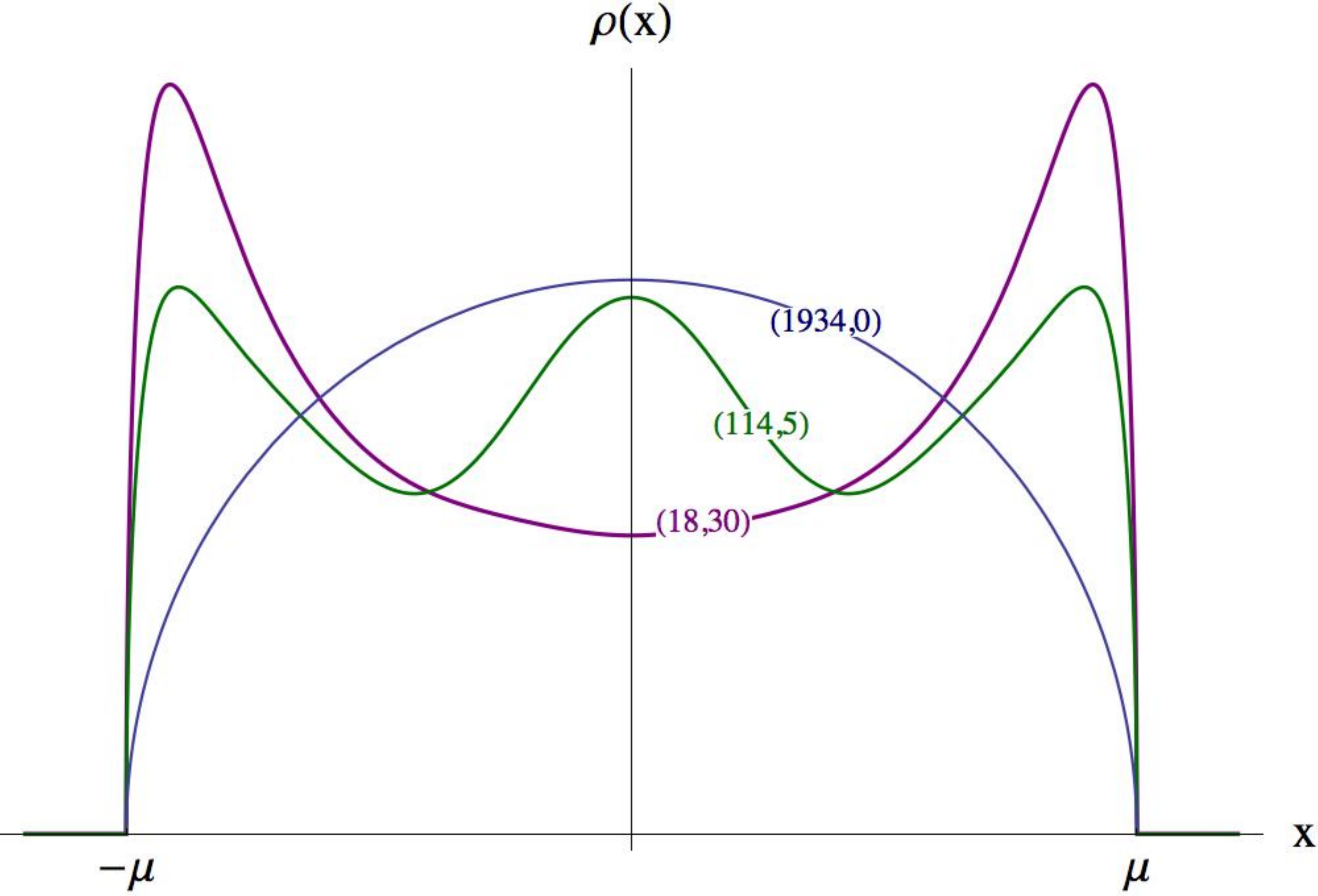}}
\caption{\label{3dens}\small The eigenvalue density for the same $\mu =7$, but at different $(\lambda ,M)$.}
\end{center}
\end{figure}
The  endpoint singularity of the eigenvalue density is of the usual square-root type across the whole phase diagram. At  small $M$ the density has a maximum at zero and monotonically decreases towards the endpoints of the distribution. But with $M$ growing, two additional maxima develop near the endpoints. In some range of parameters (for sufficiently large $\lambda $ and not so big $M$), the density has three clearly visible peaks. The peak at zero diminishes in size and at some point disappears, while the peaks near the endpoints become more and more pronounced, and asymptotically form the inverse square-root spikes of the infinite-volume density, cf.~(\ref{simpleinversesqrt}). Qualitative explanation of this behavior, which is illustrated in fig.~\ref{3dens}, is given in \cite{Russo:2012ay}.  In the large-$M$, large-$\lambda $ corner of the phase diagram, the density has a more complicated shape with many minima and maxima, due to proximity of the phase transition points in infinite volume.

\subsection{Instantons}\label{instantonsec1}

Instanton contributions are usually believed to be negligible at large $N$ because of the exponential suppression of the instanton weight\footnote{Assuming the standard 't~Hooft scaling of the gauge coupling. If the gauge coupling is kept fixed at large $N$, instantons are not suppressed \cite{Azeyanagi:2013fla}.}:
\begin{equation}\label{naive-inst}
 Z^{\rm inst}_k\sim \,{\rm e}\,^{-\frac{8\pi ^2kN}{\lambda }}.
\end{equation}
This estimate does not take into account the instanton moduli integration, which can considerably modify the instanton weight and may even overcome the exponential suppression, leading  to an instanton-induced large-$N$ phase transition \cite{Gross:1994mr}. It is reasonable to assume that all instanton contributions are either suppressed or simultaneously blow up, independently of the instanton number. It is thus sufficient to investigate the moduli space integration for a single instanton with topological charge $k=1$.

The one-instanton contribution to the partition function of the $\mathcal{N}=2^*$ theory is given by \cite{Nekrasov:2002qd,Nekrasov:2003rj}
\begin{equation}
 Z^{\rm inst}_1=-\,{\rm e}\,^{-\frac{8\pi ^2N}{\lambda }+i\theta }M^2\sum_{l=1}^{N}\prod_{j\neq l}^{}
 \frac{\left(a_l-a_j+i\right)^2-M^2}{\left(a_l-a_j\right)\left(a_l-a_j+2i\right)}\,.
\end{equation}
It also admits an integral representation:
\begin{equation}\label{intrep}
  Z^{\rm inst}_1=\,{\rm e}\,^{-\frac{8\pi ^2N}{\lambda }+i\theta }
  \frac{2M^2}{M^2+1}\int
 \frac{dz}{2\pi }\,\,
\prod_{j=1}^{N}\frac{\left(z-a_j\right)^2-M^2}{\left(z-a_j\right)^2+1},
\end{equation}
where the contour of integration encircles the poles at $a_j+i$ counterclockwise. 

At large $N$ the $z$ integral is of the saddle point type and, with exponential accuracy, 
\begin{equation}\label{zint}
  Z^{\rm inst}_1\sim \,{\rm e}\,^{-\frac{8\pi ^2N}{\lambda }}\int_{}^{}
  dz\,\,{\rm e}\,^{NS_{\rm ms}(z)}\sim 
  \,{\rm e}\,^{-NS_{\rm inst}},
\end{equation}
where the  moduli space action is
\begin{equation}
 S_{\rm ms} (z)=\int_{-\mu }^{\mu }dx\,\rho (x)\,
 \ln\frac{\left(z-x\right)^2-M^2}{\left(z-x\right)^2+1}\,,
\end{equation}
and the effective instanton action $S_{\rm inst}$ is determined by the value of $S_{\rm ms}(z)$  at the dominant saddle point:
\begin{equation}
 S_{\rm inst}=\frac{8\pi ^2}{\lambda }-S_{\rm ms}(z_*).
\end{equation}

We thus need to classify all possible saddle points, which need not lie on the contour of integration. The simplest saddle point is  $z_*=\infty $ where $S_{\rm ms}(\infty )=0$. If this saddle-point saturates the integral, the instanton action is not modified, and the naive estimate (\ref{naive-inst}) of the instanton weight holds true. This was found to happen in the conformal $\mathcal{N}=2$ SYM with $2N$ fundamental hypermultiplets \cite{Passerini:2011fe}. In the $\mathcal{N}=2^*$ case, $z_*=\infty $ is the only saddle point as long as $\mu >M$. One can look, for instance, at the behavior of the integrand on the real axis. For $|z|>M+\mu $, the integrand is exponentially small, going asymptotically to one at $z\rightarrow \infty $. Between $-M-\mu $ and $M+\mu $, the integrand rapidly oscillates, since $S_{\rm ms}(z+i0)$ has a non-zero imaginary part. The oscillatory behavior can be understood from (\ref{intrep}) as well, where the numerator changes sign each time $z$ passes through $a_j\pm M$. 
We denote the critical value of the mass, at which $\mu=M $, by $M_{c_1}$. 

If $\mu <M$, the integrand oscillates on two separate intervals $[-M-\mu ,-M +\mu ]$ and $[M-\mu ,M+\mu ]$. The gap between these intervals contains a new saddle-point at $z_*=0$. The moduli space action at this saddle-point is
\begin{equation}\label{Sinst0}
 S_{\rm ms} (0)=\int_{-\mu }^{\mu }dx\,\rho (x)\,
 \ln\frac{M^2-x^2}{1+x^2}+i\pi .
\end{equation}
Let us calculate it in the weak-coupling regime discussed in sec.~\ref{N=2*weak:sec}. The eigenvalue density then is highly peaked at zero, and the integrand can be Taylor expanded in $x$:
\begin{equation}\label{mss small l}
 S_{\rm ms}(0)= \ln M^2+i\pi-\frac{M^2+1}{M^2}\,\left\langle x^2\right\rangle +\ldots 
 = \ln M^2+i\pi-\frac{M^2+1}{M^2}\,\,\frac{\lambda }{16\pi ^2}+O\left(\lambda ^2\right).
\end{equation}
The real part of the action can be positive or negative, depending on $M$ and $\lambda $. If the action is negative, the saddle point at infinity still gives the dominant contribution. When $\mathop{\mathrm{Re}}S_{\rm ms}(0)$  changes sign the integral switches from the saddle point at $z=\infty $ to the saddle point at $z=0$. We denote the critical value of the mass by $M_{c_2}$.

\begin{figure}[t]
\begin{center}
 \centerline{\includegraphics[width=8cm]{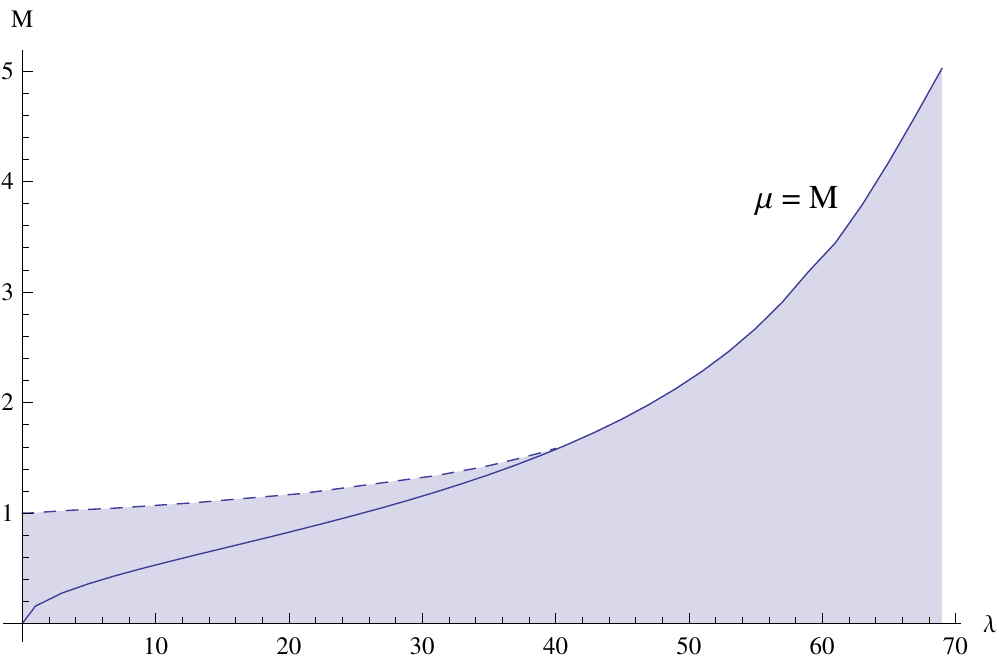}}
\caption{\label{sch}\small The phase structure of the instanton weight. The solid line is $M_{c_1}$ ($\mu =M$).
The dashed line represents $M=M_{c_2}$ and meets with $M_{c_1} $ at $\lambda \approx 40.5$. In
the shaded region the instanton action is just $8\pi ^2/\lambda $. }
\end{center}
\end{figure}
There are thus two critical lines on the $(\lambda ,M)$ plane, $M_{c_1}(\lambda )$ and $M_{c_2}(\lambda )$, at which the instanton weight discontinuously changes its behavior. At weak coupling, $\mu =\sqrt{\lambda }/2\pi $. Hence,\begin{equation}\label{Mc1}
 M_{c_1}\simeq \frac{\sqrt{\lambda} }{2\pi }\qquad \left(\lambda \rightarrow 0\right).
\end{equation}
In the other limiting case, $M\rightarrow \infty $, condition $\mu =M$ determines the second critical point $\lambda _c^{(2)}\approx 83$ that we have found in the decompactification limit (see fig.~\ref{sch2}). Consequently,
\begin{equation}
 M_{c_1}\rightarrow \infty ~~{\rm at}~~\lambda \rightarrow\lambda _c^{(2)}. 
\end{equation}
From (\ref{mss small l}) we find:
\begin{equation}\label{Mc2}
 M_{c_2}\simeq 1+\frac{\lambda }{4\pi ^2}\qquad \left(\lambda \rightarrow 0\right).
\end{equation}
The two critical lines are shown in fig.~\ref{sch}. In the shaded region on the plot the instanton action is not renormalized and the naive estimate of the instanton weight is quantitatively correct.

We thus find:
\begin{equation}
 S_{\rm inst}=\frac{8\pi ^2}{\lambda }~~{\rm if}~~ \lambda >\lambda _c^{(2)}~{\rm or}~M<\max(M_{c_1},M_{c_2}).
\end{equation}
Otherwise:
\begin{equation}
 S_{\rm inst}=\frac{8\pi ^2}{\lambda }-\int_{-\mu }^{\mu }dx\,\rho (x)\,
 \ln\frac{M^2-x^2}{1+x^2}\,.
\end{equation}
The imaginary part of $S_{\rm ms}(0)$ in the latter case  leads to renormalization of the theta-angle: $\theta \rightarrow \theta +i\pi N$.

Let us now examine  $S_{\rm inst}$ in the region where the moduli space corrections are non-trivial.  We first consider the limits where
we can compute the instanton action analytically. At weak coupling,
\be
S_{\rm inst}\simeq  \frac{8\pi ^2}{\lambda } - 2\ln M \qquad \left(\lambda \rightarrow 0\right).
\ee
Not surprisingly, at large $M$ and small $\lambda $ the moduli space corrections 
renormalize the coupling, combining into
 $8\pi ^2/\lambda _R$, where $\lambda _R$ is the running coupling of the pure $\mathcal{N}=2$ SYM.  Potentially, $\lambda _R$ can be rather big or even negative for sufficiently large $M$.
However, the approximation  (\ref{mss small l}) used in deriving this result is valid only when the {\it renormalized} coupling is small. Otherwise the eigenvalue density is no longer peaked at zero and Taylor expansion in $x$, used in deriving this result, is no longer accurate. The analysis for arbitrary $\lambda _R$ \cite{Russo:2012ay} indicates that the instanton action always remains positive, even when $\ln M$ is larger than $4\pi ^2/\lambda $.

We can also calculate the instanton action in the decompactification limit. Using (\ref{formintegrated}) we find:
\be
S_{\rm inst} =  \int_{-\mu }^{\mu }dx\,\rho (x) \ln\left(1+\frac{1}{x^2}\right)\qquad \left(M\rightarrow \infty \right).
\ee
This expression is manifestly positive. Since $\mu \propto M\rightarrow \infty $ in the decompactification limit, and the integrand is highly peaked near $x=0$, we can compute the integral as
\begin{equation}\label{rho-zero-inst}
 S_{\rm inst}= \rho (0)\int_{-\infty }^{+\infty }dx\,\ln\left(1+\frac{1}{x^2}\right)=2\pi \rho (0)\qquad \left(M\rightarrow \infty \right).
\end{equation}
The density at zero can be calculated from (\ref{densityasdisc}). Taking into account that \cite{Russo:2013qaa}
\begin{equation}
 G\left(\frac{M}{2}- i0\right)=\frac{1}{\eta }\,,\qquad 
  G\left(\frac{M}{2}+ i0\right)=\frac{1}{\bar{\eta} }\,,
\end{equation}
we get from  (\ref{densityasdisc}), (\ref{xi-eta-throughtheta}):
\begin{equation}
 S_{\rm inst}=\frac{iM\left(\eta -\bar{\eta } \right)}{\eta \bar{\eta }}
 =\frac{36i\left(\theta _4^4-\theta _3^4\right)}{M\left(E_2-2\theta _3^4+\theta _4^4\right)\left(E_2-2\theta _4^4+\theta _3^4\right)}
 \qquad \left(M\rightarrow \infty \right),
\end{equation}
where the argument of the Eisenstein series is $-r^2$ and the modulus parameter of the theta-constants is $ir$ with $r=\exp(-4\pi ^2/\lambda )$.

We verified numerically that the instanton action is positive definite throughout the whole phase diagram. We thus conclude that instantons are always exponentially suppressed in the large-$N$ $\mathcal{N}=2^*$ SYM on the four-sphere.

\section{Massive deformations of superconformal QCD}

Another way to make an $\mathcal{N}=2$ theory UV finite is to couple a vector multiplet to $2N$ hypermultiplets in the fundamental representation. When the hypermultiplets are massless this theory is superconformal and will be referred to as SCFT. The large-$N$ limit of its partition function on $S^4$ was analyzed  in \cite{Passerini:2011fe}. Here we study super-QCD-type theories obtained by relevant perturbations of $\mathcal{N}=2$ SCFT by various assignments of hypermultiplet masses. 

The localization partition function for  ${\cal N}=2$ super-QCD with an arbitrary mass assignment is given by 
\be\label{superQCD-partfunc}
Z^{\rm SQCD} =  \int d^{N-1}a\, \,\frac{\prod_{i<j}^{}\left(a_i-a_j\right)^2 H^2(a_i-a_j)}{\prod_{i,f}^{}H(a_i+M_f) H(a_i-M_f)  }
\,\,{\rm e}\,^{- \frac{8\pi^2 N} {\lambda}\sum\limits_{i}  a_i ^2}\ 
\left|{\cal Z }_{\rm inst}\right|^2.
\ee
The flavor index $f$ runs from $1$ to $N$, making the theory finite in the UV.

At large $N$, the integral is dominated by a saddle-point, determined by the equation
\be
\strokedint_{-\mu }^{\mu } dy \,\rho(y) \left(\frac{1}{x-y} -\KK(x-y)\right) 
=  \frac{8\pi^2}{\lambda}\, x - \frac{1}{2N}\sum_{f=1}^{N} \big( \KK( x + M_f)+\KK( x - M_f) \big).
\label{resa}
\ee
The qualitative structure of the solution depends on the assumptions  made about the hypermultiplet masses.
We mostly concentrate on two representative cases: (i) All masses equal, $M_f=M$. For short, we refer to this theory as ${\cal N}=2$ SCFT$^*$. (ii) Another example that we consider in detail is partially massless theory with $M_f=M$ for $f=1,\ldots ,\ N-N_0$ and $M_f=0$ for the remaining $N_0$ flavors. Later we will also consider  $\mathcal{N}=2$ SQCD with $N_f<N$ flavors, which can be obtained from the UV finite theory by making $N-N_f$ masses infinitely heavy: $M_f=\Lambda _{\rm cutoff}\rightarrow \infty $ for $f>N_f$. For simplicity we will assume that all the remaining quark masses are equal: $M_f=M$ for $f=1,\ldots ,\ N_f$.

Like  $\mathcal{N}=2^*$ theory, ${\cal N}=2$ SQCD can be viewed as a UV completion 
of pure SYM theory. The latter is obtained by making all hypermultiplets infinitely heavy, $M_f\to\infty $, and simultaneously sending $\lambda$ to zero at fixed
\begin{equation}
 \Lambda =\,{\rm e}\,^{-\frac{4\pi ^2}{\lambda }}\prod_{f=1}^{N}M_f^{\frac{1}{N}}.
\end{equation}
By expanding $\mathcal{K}(x\pm M_f)$ in (\ref{resa}) to the linear order in $x\ll M_f$, we obtain:
\be
\strokedint_{-\mu }^{\mu } dy\, \rho (y) \left(\frac{1}{x-y} -\KK(x-y)\right) 
= \frac{8\pi^2}{\lambda_{\rm r}}\,x,
\label{res2}
\ee
with the renormalized coupling defined as
\be
\frac{4\pi^2}{\lambda_{\rm r}} \equiv \frac{4\pi^2}{\lambda} -\ln \Lambda -1-\gamma.
\ee
This equation describes the large $N$ limit of pure $N=2$ Yang-Mills theory, and was studied in detail  in \cite{Russo:2012ay}.

The weak-coupling limit at fixed $M_f$ can be analyzed along the same lines as in sec.~\ref{N=2*weak:sec}, with quite similar results. We will not repeat these calculations here. The massless limit of superconformal theory was studied in \cite{Passerini:2011fe}. Here we concentrate on two other possible limits, the strong-coupling regime and  the decompactification limit. The qualitative behavior of $\mathcal{N}=2$ SQCD in these regimes is quite different from the $\mathcal{N}=2^*$ case.

\subsection{Strong coupling}

At strong coupling, $\lambda\gg 1$, the eigenvalue density extends to a large interval $(-\mu,\mu)$ which grows logarithmically with $\lambda $.
 In contradistinction to the $\mathcal{N}=2^*$ case, the eigenvalue density has a well-defined limiting shape $\rho _\infty (x)$, and one can just set $\lambda =\infty $ and $\mu=\infty $ in the saddle-point equation, to a first approximation, much like in the massless case \cite{Passerini:2011fe}. The saddle-point equation is then solved by  Fourier transform: 
\be
i \pi\mathop{\mathrm{sign}}\omega \left(1+\frac{1}{2\sinh^2 \frac{\omega}{2}}\right) \tilde \rho_\infty  (\omega )=i \pi \mathop{\mathrm{sign}}\omega
\,\frac{1}{N}\sum_{f=1}^{N}
\frac{ \cos M_f \omega }{2 \sinh^2 \frac{\omega}{2} }\,.
\ee
Hence
\be
\tilde \rho_\infty (\omega ) = \frac{ 1 }{ \cosh \omega }\,\,\frac{1}{N}
\sum_{f=1}^{N}\cos M_f\omega .
\ee
and
\be
\label{afara}
\rho_\infty (x )  = \frac{1}{4N}\sum_{f=1}^{N}\left[
 \frac {1}{\cosh\left( \frac{\pi}{2}( x+M_f)\right)}
+\frac {1}{\cosh\left( \frac{\pi}{2}( x-M_f)\right)}
\right]. 
 \ee

\begin{figure}[t]
\centering
\includegraphics[width=.5\textwidth]{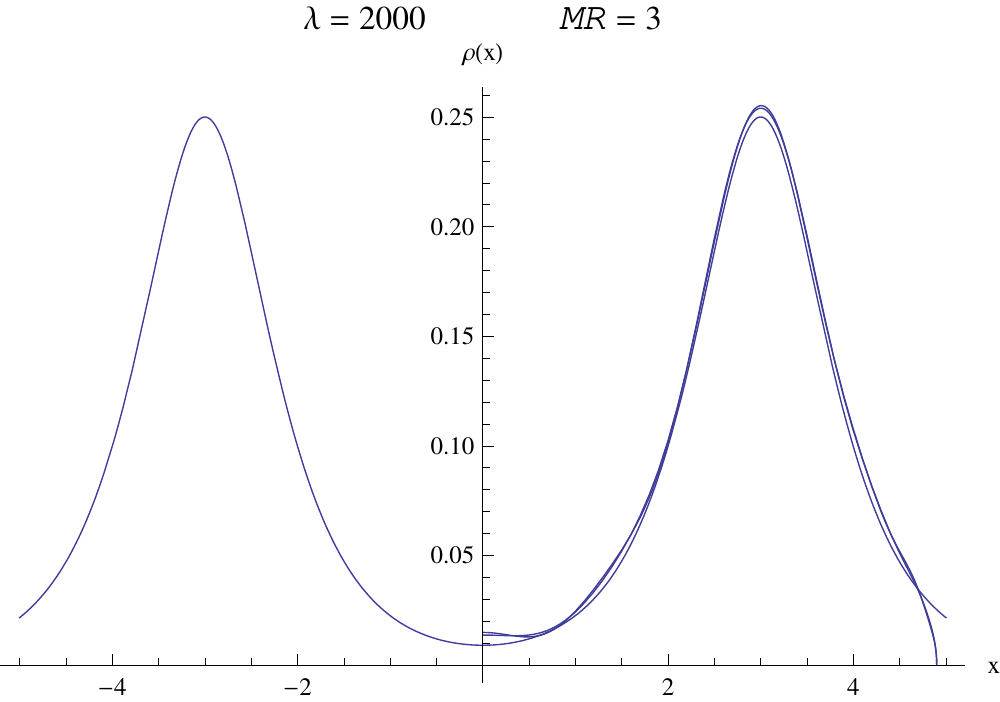} 
\caption{Eigenvalue density obtained analytically and numerically  for  ${\cal N}=2$ SYM with $2N$ fundamental hypermultiplets of equal mass at large $\lambda $. 
For clarity, the numerical eigenvalue density, which terminates at a finite value of $x$, is shown only at $x>0$. }
\label{mscf}
\end{figure}
The eigenvalue density in the ${\cal N}=2$ SCFT$^*$ (all masses equal) has two hills peaked at $x\sim \pm M$ and decays exponentially at $\pm$ infinity.
Figure~\ref{mscf} compares this analytic result with the numerical solution at $\lambda=2000$, $M=3$. For the partially massless theory, the density has three peaks, at $x\sim 0$ and $x\sim \pm M$.

The asymptotic density (\ref{afara}) has two exponential tails that extend all the way to infinity. This is not so if $\lambda $ is large but finite. The density then has to terminate at some $\mu $. This is clearly visible in the numerical solution. By matching the endpoint behavior of the density to the asymptotic solution at infinite $\lambda $, it is possible to estimate the endpoint position and the Wilson loop vev \cite{Passerini:2011fe}.
The leading order is $M$-independent:
\begin{equation}
 \mu \simeq \frac{2}{\pi }\,\ln \lambda \qquad \left(\lambda \rightarrow \infty \right).
\end{equation}
 The Wilson loop vev receives the biggest contribution from the vicinity of the endpoint, and is estimated as 
\begin{equation}\label{Wilson cube}
 W(C)\simeq \,{\rm const}\,\frac{\lambda ^3}{\left(\ln \lambda \right)^{\frac{3}{2}}}\qquad \left(\lambda \rightarrow \infty \right).
\end{equation}
Again the masses do not affect the leading-order behavior, but the constant of proportionality will depend on $M_f$. It is in principle calculable by the Wiener-Hopf method \cite{Passerini:2011fe}, but we will not attempt this calculation here.

The free energy has a finite strong-coupling limit, and can be calculated from the asymptotic solution (\ref{afara}). We have
\begin{eqnarray}
 \frac{\partial F}{\partial M_f}&=&\frac{1}{N}\left\langle 
 \mathcal{K}\left(x-M_f\right)- \mathcal{K}\left(x+M_f\right)
 \right\rangle
 \\
 \label{dF-SQCD}
 \frac{\partial F}{\partial\lambda}& =& -\frac{8\pi^2}{\lambda^2} \langle x^2\rangle\ .
\end{eqnarray}
Using the eigenvalue density (\ref{afara}) we find
 \be
\langle x^2\rangle_\infty =1+\frac{1}{N}\sum_{f}^{}M^2_f
\ee
and
\begin{equation}
 \left\langle \mathcal{K}(x+m)\right\rangle
 =-\frac{1}{2N}\sum_{f}^{}\frac{\partial }{\partial m}\left(
 \ln\mathcal{F}\left(m+M_f\right)+\ln\mathcal{F}\left(m-M_f\right)
 \right), 
\end{equation}
where the function $\mathcal{F}(x)$ is defined as
\begin{equation}
 \mathcal{F}(x)=H(x)\,\frac{\Gamma \left(\frac{1+ix}{4}\right)\Gamma \left(\frac{1-ix}{4}\right)}{\Gamma \left(\frac{3+ix}{4}\right)\Gamma \left(\frac{3-ix}{4}\right)}
\end{equation}
Hence
\begin{equation}
 N^2F^{\rm SQCD}=\ln\left(\prod_{ff'}^{}\mathcal{F}\left(M_f+M_{f'}\right)\mathcal{F}\left(M_f-M_{f'}\right)\right)
+\frac{8\pi ^2N}{\lambda }\sum_{f}^{}\left(1+M_f^2\right)+O\left(\frac{1}{\lambda^2 }\right).
\end{equation}
For ${\cal N}=2$ SCFT$^*$ this gives, up to a constant:
\begin{equation}
 F^{ {\rm SCFT}^*}=\ln \mathcal{F}(2M)+\frac{8\pi ^2(1+M^2)}{\lambda }+O\left(\frac{1}{\lambda^2 }\right).
\end{equation}

\subsection{Decompactification}

The decompactification limit can be analyzed much in the same way as  in sec.~\ref{decompact:sec}. First we recover the dependence on $R$ by rescaling all dimensionful quantities and then send $R$ to infinity. Once the resulting equation is differentiated twice, the kernel becomes algebraic, because $\mathcal{K}(x)$ can be replaced by its large-argument asymptotics (\ref{KKapprox}). We will analyze separately two special cases, the $\mathcal{N}=2$ SCFT$^*$  with equal hypermultiplet masses and partially massless theory. 

\subsubsection{SCFT$^*$}

In the $\mathcal{N}=2$ SCFT$^*$ case, the steps described above lead to the following simple equation:
\begin{equation}
\label{inta}
 2\strokedint_{-\mu }^\mu dy\,\,
 \frac{\rho (y)}{x-y}=\frac{1}{M+x}-\frac{1}{M-x}\,.
\end{equation}
This looks like the saddle-point equation for a one-matrix model with a logarithmic potential. Slightly more general matrix model with an additional quartic potential
was considered in  \cite{Kazakov:1989cq}, as a model
for open strings in zero dimensions. The model has a rich phase structure and exhibits quite non-trivial critical behavior, governed by an interplay between the logarithmic and polynomial terms in the potential\footnote{We would like to thank V.~Kazakov for comments on this point.}. More general mass assignment in SQCD can probably mimic additional terms in the effective matrix-model potential and thus can lead to an interesting critical behavior.

In our case, the effective potential is actually upside-down, which should not worry us too much, as the boundary conditions here are different compared to usual matrix models.  In the matrix model language, we need to find the solution squeezed between two infinite walls at $x=\pm \mu $. The unique normalizable solution with such boundary conditions exists for any $\mu $. In contradistinction to the usual matrix models, normalization does not fix the endpoint positions $\pm\mu $, which are rather determined by the integrated form of (\ref{inta}), equivalent to the original saddle-point equation after the first differentiation:
\be
 \int_{-\mu }^{\mu }dy\,\rho (y)\ln \frac{y^2}{M^2} =-\frac{8\pi ^2}{\lambda }\,.
\label{orsad}
\end{equation}

The unique normalizable solution of  (\ref{inta}), at fixed $\mu $, is given by 
\be
\rho(x)=\frac{M\sqrt{M^2-\mu^2}}{\pi}\ \frac{1}{ \sqrt{\mu^2-x^2}}\, \frac{1}{M^2-x^2}\ .
\ee
In order to find $\mu $ as a function of $\lambda $ and $M$, we substitute $\rho $ into
(\ref{orsad}).
Using 
\be
\label{ottt}
\left\langle \ln\frac{x^2}{M^2}\right\rangle=2\ln\frac{\mu}{M+\sqrt{M^2-{\mu^2}}}\ ,
\ee
we obtain
\be\label{M-SCF*}
\mu=\frac{M}{\cosh  \frac{4\pi^2}{\lambda} }\,.
\ee

Note that $\mu$ never exceeds $M$. As a result there are no phase transitions in this model. The reason can be easily understood from the saddle-point equation  (\ref{inta}). The effective potential in the analog matrix model is unbounded from below and as soon as $\mu $ approaches $M$ the eigenvalues start to fall down the infinite potential well. The attractive force acting towards $x=\pm M$ becomes stronger and stronger when $\mu $ approaches $M$ and can overcome mutual repulsion between the eigenvalues, compressing larger and larger number of them towards the endpoint of the distribution. A natural question is how to  reconcile this behavior
with the strong-coupling behavior studied in the previous section.
In the latter case, $\mu $ grows like $\ln \lambda $ and can certainly exceed $M$.
However, the eigenvalues sitting at $|x|>M$ represent the exponential tail of the distribution which
vanishes as $M\to \infty$. 
This tail is automatically cutoff  when the limit $M=\infty $ is taken before considering $\lambda\gg 1$, and in this case the eigenvalue distribution
always ends at $\mu<M$. In fact, in the limit where both $M\to\infty $ and $\lambda\to \infty$, taken in any order, the eigenvalue density approaches two delta functions  peaked at
$x=\pm M$.

The free energy can be found from (\ref{dF-SQCD}). For the second moment of the eigenvalue density we get:
\begin{equation}
 \left\langle x^2\right\rangle=M^2-M\sqrt{M^2-\mu^2}=
 M^2\left(1-\tanh\frac{4\pi ^2}{\lambda }\right),
\end{equation}
which gives:
\be\label{freeen-SCF*}
F=- 2M^2\ln \left(1+\,{\rm e}\,^{-\frac{8 \pi ^2}{\lambda }}\right)+{\rm const}.
\ee
Strikingly, the free energy of the $\mathcal{N}=2$ SCFT$^*$ is given by the first $n=1$ term of the free energy (\ref{freefe})  of 
${\cal N}=2^*$ SYM.

The weak-coupling expansion of (\ref{M-SCF*}), (\ref{freeen-SCF*}) has the expected OPE form (\ref{weak-expansion}). For instance,
\be
\label{frese}
F = 2M^2\sum_{k=1}^\infty {(-1)^k\over k}\,\,{\rm e}\,^{-\frac{8 \pi ^2k}{\lambda }}.
\ee
The simplicity of the expansion coefficients again suggests that there may be a more direct way to calculate them, without the use of localization.

Computing the circular Wilson loop, we find:
\begin{equation}\label{wilp}
 W(C)=\left\langle \,{\rm e}\,^{2\pi Rx}\right\rangle
 \simeq 
 \frac{1}{2\pi\sqrt{ M R}}\frac{\cosh^{3\over2} \frac{4\pi^2}{\lambda}  }{\sinh\frac{4\pi^2}{\lambda}} \exp \left(\frac{2\pi MR}{\cosh \frac{4\pi^2}{\lambda} }\right).
\end{equation}
Since $R\rightarrow \infty $,  the main contribution comes from the region near $x\approx \mu $.  We thus conclude that large Wilson loops obey  perimeter law with the coefficient given by $\mu $ in (\ref{M-SCF*}). At strong coupling, the coefficient just asymptotes to $M$. The prefactor, as a function of the coupling constant, grows linearly at large $\lambda $. This is different from the result (\ref{Wilson cube}), found by  taking the limit $\lambda \rightarrow \infty $ first. The discrepancy is not surprising, since the Wilson loop is sensitive to the exponential tail of the eigenvalue density, which is cut off in the $MR=\infty$ limit that we are considering now.

An interesting feature of the Wilson loop vev (\ref{wilp}) is that the exponent does not have a $\sqrt{\lambda}$ coefficient as one would expect from a string world-sheet interpretation.
Recall that, for ${\cal N}=4$ SYM,  the coefficient $\sqrt{\lambda}$ in the exponent arises from the squared radius  of AdS space in units of $\alpha '$.
Likewise, perimeter law in the $\mathcal{N}=2^*$ SYM at strong coupling bears a factor of $\sqrt{\lambda }$, with the coefficient in exact agreement with the area law in the geometry of the holographic dual \cite{Buchel:2013id}. Strings in the
supergravity dual of massive ${\cal N}=2$ SCFT$^*$, which is not known, are  likely to have quantum and  highly interacting worldsheet even in the limit of large 't~Hooft coupling. 

\subsubsection{Partially massless theory}

If $N_0$ flavors are left massless, double differentiation of the saddle-point equation leads to 
\begin{equation}
\label{intaa}
 2\strokedint_{-\mu }^\mu dy\,\,
 \frac{\rho (y)}{x-y}= \frac{\nu }{M+x}  -\frac{\nu }{M-x}+  \frac{2\left(1-\nu \right)}{x}\,,
\end{equation}
where  $\nu $ denotes the fraction of massive flavors:
\begin{equation}
 \nu=\frac{N-N_0}{N}\,.
\end{equation}
In the large-$N$ limit, $\nu $ is a real number between zero and one. The force term on the right-hand side now has a singularity on the interval $(-\mu ,\mu )$. The inversion of the Hilbert kernel thus becomes ambiguous, and depends on  how the singularity is regularized. The limiting procedure ($R\rightarrow \infty $), that was used in deriving (\ref{intaa}), actually dictates a very concrete regularization prescription. Indeed, the $1/x$ driving term on the right-hand side arises from approximating  $\mathcal{K}(xR)$  by $xR\log|xR|$ in  the limit $R\rightarrow \infty $, eq.~(\ref{KKapprox}). This approximation is clearly inapplicable when $x\rightarrow 0$. In fact, before the limit was taken, the original function $\mathcal{K}(xR)$ had been non-singular at zero. Consequently, $1/x$ is smoothened out on the scale $\epsilon \sim 1/R$. Since $\mathcal{K}(x)$ is an odd function, the smoothened $1/x$ will automatically retain anti-symmetry under $x\rightarrow -x$. Such regularization is equivalent to the principal-value prescription
$$
 \frac{2}{x}~\longrightarrow ~\frac{1}{x+i\epsilon }+\frac{1}{x-i\epsilon }\,.
$$
If $1/x$ is understood in this way, the normalizable solution to (\ref{intaa}) is given by
\be
\label{rhonn}
\rho(y)=\frac{\nu M\sqrt{M^2-\mu^2}}{\pi}\,\,\frac{1}{ \sqrt{\mu^2-x^2}}\,\, \frac{1}{M^2-x^2}+ (1-\nu )\delta(x).
\ee

The analog of (\ref{orsad}) now reads\footnote{Eq.~(\ref{intaa}) guarantees that the right-hand-side is constant independent of $x$. In (\ref{orsad}) we have thus set $x=0$ without loosing any information. Here we cannot set $x=0$ directly, because of the log singularity, and thus prefer to keep $x$ as a parameter. }
\be
 \int_{-\mu }^{\mu }dy\,\rho (y)\ln\left(x-y\right)^2
-\nu  \ln\left(M^2-x^2\right)-(1- \nu ) \ln x^2
 =-\frac{8\pi ^2}{\lambda }\,.
\label{orsadd}
\end{equation}
Evaluating the left-hand side on the solution (\ref{rhonn}), we find:
\begin{equation}
\mu = \frac{M}{\cosh \frac{4\pi^2 }{\lambda \nu }}\,.
\ee
The only effect of the massless multiplets, compared to $\mathcal{N}=2$ SCFT$^*$, is a rescaling of $\lambda $.
In particular, $\mu$ still obeys the bound $\mu<M$. 
The Wilson loop is given by the same formula as in the previous section with $\lambda $ rescaled. Likewise, for the free energy we get:
\be\label{partmasslessfree}
F=- 2M^2\nu ^2\ln \left(1+\,{\rm e}\,^{-\frac{8 \pi ^2}{\lambda \nu  }}\right)+{\rm const}.
\ee

The OPE expansion now goes in powers of $\exp(-8 \pi ^2/\lambda \nu )$. This is easy to understand. The low-energy sector of the model, left upon integrating out massive fields, is  $\mathcal{N}=2$ SQCD with $2N_0$ massless hypermultiplets. The beta functions of this theory is
\begin{equation}
 \beta _{\rm SQCD}=-\frac{g_{\rm YM}^4N(N-N_0) }{4\pi ^2}=-\frac{\lambda ^2\nu }{4\pi ^2}\,,
\end{equation}
and the dynamically generated scale is given by
\begin{equation}
 \Lambda _{\rm SQCD}=M\,{\rm e}\,^{-\frac{4\pi ^2}{\lambda \nu }}.
\ee
The OPE goes in powers of  $\Lambda^2_{\rm SQCD}/M^2$, which now translates into $\exp(-8 \pi ^2/\lambda \nu )$.

\subsection{Instantons}\label{instantonsec2}

The one-instanton  contribution to the partition function of the $\mathcal{N}=2$ $SU(N)$ super Yang-Mills theory 
with $2N$ hypermultiplets can be obtained from the general formulas given in \cite{Nekrasov:2002qd,Nekrasov:2003rj}:
\begin{equation}
 Z^{\rm inst}_1= {\rm e}\,^{-\frac{8\pi ^2N}{\lambda }+i\theta }  \sum_{l=1}^{N}
 \frac{\prod\limits_{f}^{}\left[\left(a_l+i\right)^2-M_f^2\right]}{\prod\limits_{j\neq l}^{}\left(a_l-a_j\right)\left(a_l-a_j+2i\right)}\,,
\end{equation}
and has an integral representation:
\begin{equation}\label{instM}
  Z^{\rm inst}_1=2\,{\rm e}\,^{-\frac{8\pi ^2N}{\lambda }+i\theta }
 \int
 \frac{dz}{2\pi }\,\,
\frac{\prod\limits_{f}^{}\left(z^2-M_f^2\right)}{\prod\limits_{j}\left[\left(z-a_j\right)^2+1\right]} \, ,
\end{equation}
where the contour of integration encircles the poles at $a_j+i$ counterclockwise.

The large-$N$ limit of the instanton contribution was examined  in \cite{Passerini:2011fe} for $M_f=0$. It was found that the exponential part of the instanton weight is not modified by the moduli integration.
We extend this analysis to the massive SCFT$^*$ with equal hypermultiplet masses.

The moduli-space action,
\begin{equation}
 S_{\rm ms}(z)=\ln\left(z^2-M^2\right)-\int_{-\mu }^{+\mu }dx\,\rho (x)\ln\left(\left(z-x\right)^2+1\right),
\end{equation}
has two saddle-points at $z=0$ and $z=\infty $. Since $S_{\rm ms}(\infty )=0$, the instanton action is not renormalized if the dominant saddle-point is the one at infinity. At zero,
\begin{equation}
 S_{\rm ms}(0)=\left\langle \ln\frac{M^2}{x^2+1}\right\rangle-i\pi .
\end{equation}
For the asymptotic solution at strong coupling, eq.~(\ref{afara}), the real part of the  saddle-point action is always negative:
\begin{equation}
 \left\langle \ln\frac{M^2}{x^2+1}\right\rangle_\infty =2
 \ln\tanh\frac{\pi M}{4}<0.
\end{equation}
We thus conclude that at strong coupling the trivial saddle-point at infinity is dominant.

In the decompactification limit, 
\begin{equation}
 \left\langle \ln\frac{M^2}{x^2+1}\right\rangle=\left\langle \ln\frac{x^2}{x^2+1}\right\rangle+\frac{8\pi ^2}{\lambda }\qquad \left(M\rightarrow \infty \right),
\end{equation}
in virtue of (\ref{orsad}). This expression is positive, so the saddle-point at $z=0$ is dominant. The same steps that led to (\ref{rho-zero-inst}), now give for the instanton action
\begin{equation}
 S_{\rm inst}=2\pi \rho (0)=\frac{2}{M}\,\sinh\frac{4\pi ^2}{\lambda }\,.
\end{equation}

There is a line $M_c(\lambda )$ in the $(\lambda ,M)$ plane below which the suppression is determined solely by the instanton action $8\pi ^2/\lambda $. In the region above $M_c$ the instanton action is renormalized by the moduli-space integration. As in sec.~\ref{instantonsec1}\footnote{$M_c$ of the SCFT$^*$ is analogous to $M_{c_2}$ of $\mathcal{N}=2^*$ SYM.}, $M_c(0)=1$. At $\lambda \rightarrow \infty $, $M_c$  goes to infinity. 
We have computed $S_{\rm inst}$ numerically in the region above $M_c$ and found that it is always positive definite. 
In conclusion, the one-instanton contribution is exponentially suppressed in the large-$N$ limit.

\section{Super-QCD with $2N_f$ massive hypermultiplets}

Another interesting  theory  is 
  ${\cal N}=2$ SQCD, $\mathcal{N}=2$ supersymmetric gauge theory with  
   $2N_f$ massive hypermultiplets of equal mass $M$. We would like to study this theory in  the Veneziano limit, $N\rightarrow \infty $, $N_f\rightarrow \infty $ with $N_f/N$ fixed \cite{Veneziano:1976wm}.
 We shall assume that $N_f<N$, in which case the theory is asymptotically free, and interpolates between pure ${\cal N}=2$ SYM at $N_f=0$ and superconformal YM at $N_f=N$. 
 
 A neat way to define the partition function of $\mathcal{N}=2$ SQCD is to start with a UV finite partition function (\ref{superQCD-partfunc}) and make $N-N_f$ quarks infinitely heavy, while keeping the mass of the remaining $N_f$ quarks fixed.  The mass of the heavy flavors, which we denote by $\Lambda _0$, serves as a UV cutoff. Using (\ref{Klarge}),
 $$
\ln H\left(a+\Lambda _0\right)+\ln H(a-\Lambda _0)\simeq 
2\ln H(\Lambda _0)- 2\left(\ln\Lambda _0+\gamma +1\right)a^2
$$
the contribution of the heavy fields can be absorbed in the renormalization of the 't~Hooft coupling:
\be
\frac{4\pi ^2 N}{\lambda_R} = \frac{4\pi ^2 N}{\lambda} - (N-N_f)\left(\ln \Lambda _0 +\gamma+1\right).
\ee
Introducing 
the Veneziano parameter
\begin{equation}
\zeta =\frac{N_f}{N}\,,
\end{equation}
and the dynamically generated scale
\begin{equation}
 \Lambda =\Lambda _0\,{\rm e}\,^{-\frac{4\pi ^2}{\lambda \left(1-\zeta \right)}},
\end{equation}
the partition function can be written as 
\be\label{NfQCD-partfunc}
Z^{\rm SQCD}_{N_f} =  \int d^{N-1}a\, \,\frac{\prod_{i<j}^{}\left(a_i-a_j\right)^2 H^2(a_i-a_j)}{\prod_{i}^{}H^{N_f}(a_i+M) H^{N_f}(a_i-M)  }
\,\,{\rm e}\,^{2(N-N_f)\left(\ln\Lambda +\gamma +1\right)\sum\limits_{i}  a_i ^2}\ 
\left|{\cal Z }_{\rm inst}\right|^2.
\ee

The saddle-point equation takes the following form:
\be
2\strokedint_{-\mu }^{\mu } dy \rho(y) \left(\frac{1}{x-y} -\KK(x-y)\right) 
= -4\left(1-\zeta \right) \left(\ln \Lambda+\gamma +1\right)x - \zeta   \KK( x + M) - \zeta   \KK( x - M).
\label{fresa}
\ee
The  model depends on
two parameters $(\Lambda, M)$, and we may consider several limiting cases in which the saddle-point equation simplifies and can be explicitly solved. For instance, if $M\gg \Lambda $, the hypermultiplets can be integrated out leaving behind pure $\mathcal{N}=2$ SYM with
\begin{equation}\label{Leff-SQCD}
 \Lambda _{\rm eff}=\Lambda ^{1-\zeta }M^\zeta .
\end{equation}
In the perturbative regime  of a very small sphere ($\Lambda\ll 1,\,M\lesssim 1 $), the linear force is strong and attractive and pushes all eigenvalues towards the origin. The solution is Wigner's semicircle (\ref{WignerLaw}) with $\mu =\sqrt{\lambda _R}/2\pi $. 

A more interesting case is the opposite, decompactification regime when $\Lambda\gg 1,\,M\gg 1 $. 
 In the decompactification limit the linear force is strong but repulsive. Eigenvalues are pushed away from the origin and extend over a large
interval $(-\mu,\mu)$. As a result, $x,y$ are on average large. One can approximate the $\KK $ function by its asymptotic form (\ref{KKapprox}). 
Differentiating the saddle-point equation twice we get:
\be
\label{mfre}
2\strokedint_{-\mu }^{\mu } dy\,\,  \frac{\rho(y)}{x-y}
=   \frac{\zeta }{x+M} +\frac{\zeta }{x-M}\,.
\ee
This is a very simple equation, but it exhibits the same phenomenon as we encountered in the decompactification limit of the $\mathcal{N}=2^*$ theory \footnote{Again, the equation is the same as in the matrix model of \cite{Kazakov:1989cq}, but with different boundary conditions.}. Namely, the driving term has poles at $x=\pm M$ which may or may not lie within the eigenvalue distribution. The poles are associated with massless hypermultiplets which appear in the spectrum as soon as the poles cross the boundary of the eigenvalue distribution. The model thus has two phases: the weak-coupling phase with $\mu <M$, in which all hypermultiplets are heavy, and the strong-coupling phase at $\mu >M$, in parametrically light hypermultiplets appear in the spectrum.

We begin with the strong-coupling phase ($\mu >M$). The normalized eigenvalue density is then given by
\be
\label{noj}
\rho (x) = \frac{1-\zeta }{\pi \sqrt{\mu^2-x^2}} + \frac{\zeta }{2}\,\delta(x+M)+\frac{\zeta }{2}\,\delta(x-M).
\ee
To find $\mu $, we use the integrated form of (\ref{mfre}):
\begin{equation}\label{auxiliary-SQCD}
 \int_{-\mu }^{+\mu }dy\,\rho (y)\ln\frac{y^2}{M^2}=\left(1-\zeta \right)\ln\frac{\Lambda ^2}{M^2}\,.
\end{equation}
 Substituting the solution (\ref{noj}) we obtain:
\begin{equation}\label{mu=2Lambda}
 \mu =2\Lambda .
\end{equation}
This generalizes the solution of the pure $\mathcal{N}=2$ SYM  in the 't~Hooft limit \cite{Douglas:1995nw,Ferrari:2001mg,Russo:2012ay}$^3$ to the solution of  $\mathcal{N}=2$ SQCD in the Veneziano limit. Interestingly, the width of the eigenvalue distribution does not depend on the hypermultiplet mass.

The solution was obtained under the assumption that the width of the eigenvalue distribution exceeds the hypermultiplet mass: $\mu >M$. This condition breaks down when $M$ reaches $\mu =2\Lambda $. As a result, the system undergoes a transition to the weak-coupling regime. The phase transition thus happens at 
\begin{equation}
 M_c=2\Lambda .
\end{equation}

We now proceed with solving the model in the phase with $\mu <M$. The saddle-point equation (\ref{mfre}) is then very similar to (\ref{inta}). The  solution defined on the interval $(-\mu ,\mu )$ and having unit normalization is given by
\begin{equation}
 \rho (x)=\frac{1}{\pi \sqrt{\mu ^2-x^2}}\left(1-\zeta +\frac{\zeta M\sqrt{M^2-\mu ^2}}{M^2-x^2}
 \right).
\end{equation}
To find $\mu $ we substitute the density into (\ref{auxiliary-SQCD}) and use (\ref{ottt}):
\begin{equation}
 \left\langle \ln\frac{x^2}{M^2}\right\rangle=2\left(1-\zeta \right)\ln\frac{\mu }{2M}+2\zeta \ln\frac{\mu }{M+\sqrt{M^2-\mu ^2}}\,.
\end{equation}
This results in a transcendental equation for $\mu $, whose solution can be written in a parametric form:
\begin{eqnarray}\label{muuu}
\mu &=&M\sqrt{1-u^2},
\\
 \left(\frac{2\Lambda }{M}\right)^{2-2\zeta }&=&\left(1+u\right)^{1-2\zeta }\left(1-u\right).
 \label{udefinition}
\end{eqnarray}
With increasing $M$, $ \mu $ grows from $\mu =2\Lambda $ at the critical point  to infinity, asymptotically approaching $$\mu \simeq 2\Lambda ^{1-\zeta }M^\zeta=2\Lambda _{\rm eff}, $$
in accord with effective-field theory expectations. Indeed, at very large $M$ the model should be equivalent to pure $\mathcal{N}=2$ SYM with the effective scale (\ref{Leff-SQCD}).

As we have shown, the model undergoes a phase transition at $M=2\Lambda $.
To determine the order of the  transition we can calculate the free energy, or better its first derivative:
\begin{equation}
 \frac{\partial F}{\partial \ln\Lambda }=-2\left(1-\zeta \right)\left\langle x^2\right\rangle.
\end{equation}
 For the second moment of the density we get:
\begin{equation}\label{<x^2>-strong}
 \left\langle x^2\right\rangle=2\left(1-\zeta \right)\Lambda ^2+\zeta M^2\qquad \left(M<2\Lambda \right),
\end{equation}
in the strong-coupling phase, and
\begin{equation}
 \left\langle x^2\right\rangle=\frac{M^2}{2}\left(1-u\right)\left[1+\zeta +\left(1-\zeta \right)u\right]\qquad \left(M>2\Lambda \right),
\end{equation}
in the weak-coupling phase. To compare the two expressions it is convenient to rewrite the strong-coupling answer (\ref{<x^2>-strong}) in terms of the variable $u$ defined in (\ref{udefinition}):
\begin{equation}
 \left\langle x^2\right\rangle=\frac{M^2}{2}\left[2\zeta +\left(1-\zeta \right)\left(1+u\right)^{\frac{1-2\zeta }{1-\zeta }}\left(1-u\right)^{\frac{1}{1-\zeta }}\right] \qquad \left(M<2\Lambda \right).
\end{equation}
The phase transition happens at $u=0$. Taylor expanding around the critical point, we find:
\begin{equation}
 \left\langle x^2\right\rangle=\frac{M^2}{2}\times 
\begin{cases}
 1+\zeta -2\zeta u-\left(1-\zeta \right)u^2 & {\rm }u>0
\\
 1+\zeta -2\zeta u-\frac{1-2\zeta -\zeta ^2}{1-\zeta }\,u^2+\ldots  & {\rm }u<0
\end{cases}
\end{equation}
The first two Taylor coefficients coincide. Consequently, the free energy is continuous up to the third derivative, which has a finite jump at the transition point. The phase transition is thus of the third order.

\begin{figure}[t]
\begin{center}
 \centerline{\includegraphics[width=8cm]{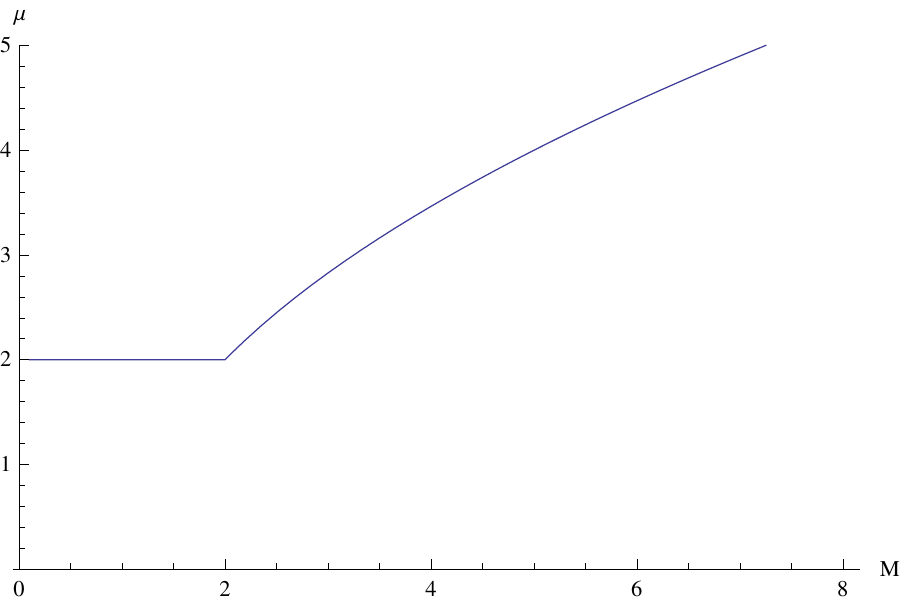}}
\caption{\label{mu-vs-M}\small The width of the eigenvalue distribution as a function of the quark mass at fixed $\Lambda $ and $\zeta =1/2$ ($N=2N_f$). The units on both axes are normalized to $\Lambda $.}
\end{center}
\end{figure}
The Wilson loop satisfies perimeter law with the exponent dictated by  $\mu $:
\begin{equation}
 \ln W(C)\simeq \mu L
\end{equation}
for a contour\footnote{Using localization we can compute the circular Wilson loop on the sphere, for which the exponent is equal to $2\pi \mu  R$ and length is equal to $2\pi R$. We extrapolate this perimeter-law behavior to any sufficiently large contour.} of length $L$. The coefficient of proportionality is given by (\ref{mu=2Lambda}) at $M<2\Lambda $ and by (\ref{muuu}), (\ref{udefinition}) at $M>2\Lambda $. The dependence of $\mu $ on $M$ is plotted in fig.~\ref{mu-vs-M}. It is clear from the plot, and also from eqs.~(\ref{mu=2Lambda}), (\ref{muuu}), (\ref{udefinition}), that $\mu $ is continuous across the phase transition but its first derivative experiences a jump. 

The plot in fig.~\ref{mu-vs-M} is for  $\zeta  =1/2$. In this particular case one can express $\mu $ in terms of $M$ and $\Lambda $ explicitly:
\begin{equation}
\mu^2 = 4 \Lambda (M- \Lambda )\qquad \left(\zeta =1/2,~M>2\Lambda\right)
\end{equation}
The second moment of the eigenvalue density takes the form:
\be
\langle x^2\rangle = \Lambda^2+ \frac{M^2}{2}\qquad (\zeta =1/2,~M<2\Lambda)
\ee
and
\be
\langle x^2\rangle = \Lambda (2M-\Lambda )\qquad (\zeta =1/2,~M> 2\Lambda).
\ee
Its second derivative is discontinuous at the transition point.

The phase transition in $\mathcal{N}=2$ SQCD shares a lot of similarities with  the phase transition found in the ${\cal N}=2^*$ theory, although the SQCD case is technically much simpler.
In both cases the phase transition is caused by the emergence of massless modes, which produce resonance peaks in the eigenvalue density.
In the present case, these peaks are delta functions $\delta(x\pm M)$. As a result, the transition is less continuous than in the  ${\cal N}=2^*$ case.

\section{Conclusions}

Massive $\mathcal{N}=2$ gauge theories exhibit a wealth of non-perturbative phenomena in the large-$N$ limit. In specific limits the physics is described by solvable matrix models.
We have found that theories with dimensionless couplings, such as $\mathcal{N}=2^*$ SYM, or with two mass scales, such as SQCD, undergo large-$N$ phase transition as the couplings or mass ratios change. 
On the other hand, we found that $SU(N)$ super-Yang-Mills
with $2N$ massive hypermultiplets in the fundamental exhibits a continuous interpolation between the weak and strong coupling regimes, without phase transitions.
The models studied in this paper have the expected OPE expansion, albeit with rather simple coefficients. This fact is probably due to supersymmetry.
We have also given explicit formulas for the 1/2 BPS circular Wilson loop and free energy in the strong coupling limit, for the different models. This may allow for 
a direct comparison with formulas obtained from holographic dual candidates.

When phase transitions occur, the weak and strong coupling regimes  correspond to different, disconnected branches of the large-$N$ master field. This could potentially pose a problem for the holographic description, for instance in the context of the $\mathcal{N}=2^*$ theory, where an infinite number of phase transitions accumulate at strong coupling. However, the mere existence of phase transitions does not preclude the string description from being exactly equivalent to field theory. The phase transitions should be then visible on the string-theory side as well. We have no clear idea what mechanism can trigger phase transitions in the string sigma-model, but presumably they are related to the singularities that the supergravity background \cite{Pilch:2000ue,Buchel:2000cn} in which the string propagates has in the far IR.

The localization result of \cite{Pestun:2007rz} is a plug-in formula valid in principle for any $\mathcal{N}=2$ theory on $S^4$. Its generalization to the squashed four-sphere is also known \cite{Hama:2012bg,Fucito:2013fba}. It would be interesting to investigate the large-$N$ limit of $\mathcal{N}=2$ theories in both cases in more generality, or at least to go through a larger set of examples. As a first step in this direction we briefly comment on the large-$N$ limit of certain quiver models in appendix~\ref{quiver:appendix}.

\subsection*{Acknowledgments}

We would like to thank N.~Bobev, A.~Buchel, N.~Drukker, G.~Festuccia, N.~Gromov, V.~Kazakov, I.~Kostov, Y.~Makeenko, K.~Skenderis and D.~Volin for discussions.
The work of K.Z. was supported in part by People Programme (Marie Curie Actions) of the European Union's FP7 Programme under REA Grant Agreement No 317089. J.R. acknowledges support by MCYT Research
Grant No.  FPA 2010-20807.


\appendix

\section{Functions $H(x)$ and $\KK(x)$}\label{kernelappendix}

Here we collect some formulas for the functions $H(x)$ and $\KK(x)$, defined in (\ref{functionH}) and (\ref{functionKK}), respectively. The former is related to the Barnes $G$-function:
\be
H(x)= e^{-(1+\gamma)x^2} G(1+ix)G(1-ix).
\ee
The latter can be expressed through the $\psi $-function (the logarithmic derivative of the $\Gamma$-function):
\be
\KK(x) =x \left( \psi(1+ix) +\psi(1-ix)+2\gamma \right),
\ee
where $\gamma=-\psi (1) $ is the Euler constant.

The function  $\mathcal{K}(x)$ is meromorphic on the whole complex plane, and has an infinite series of poles along the imaginary axis. The residue at the $n$th pole grows linearly with $n$.
It can be viewed as a generating function of odd zeta-values: 
\be\label{zeta-expansion}
\KK(x) = 2 \sum_{k=1}^\infty (-1)^{k+1} \zeta(2k+1) x^{2k+1}.
\ee
The expansion is convergent for $|x|<1$, until the first pole of the $\psi $-function at $x=\pm i$. 

At large real values of the argument, the functions $H(x)$ and $\mathcal{K}(x)$ behave as
\bea\label{Klarge}
\ln H(x)&=&-\frac{1}{2}\,x^2\ln x^2+\left(\frac{1}{2}-\gamma \right)x^2+O\left(\ln x^2\right)\\
\label{KKapprox}
\KK(x)&=&x\ln x^2+2\gamma x+O(x^{-1}).
\eea
These are the first terms of an asymptotic expansion with zero radius of convergence.

Another useful formula is the Fourier-transform representation of $\mathcal{K}''(x)$:
\begin{equation}\label{FourierKprpr}
 \mathcal{K}''(x)=4\int_{0}^{\infty }d\omega \,\,
 \frac{\omega ^2\sin 2\omega x}{\sinh^2\omega }\,.
\end{equation}

\section{Theta functions and Eisenstein series}\label{thetappendix}

Here we list some functions that appear in the  exact solution of the $\mathcal{N}=2^*$ matrix model. Those are the four theta-functions:
\begin{eqnarray}
 \theta _1(z|q)&=&2q^{\frac{1}{4}}\sum_{n=0}^{\infty }\left(-1\right)^n
 q^{n\left(n+1\right)}\sin\left(2n+1\right)z
\nonumber \\
\theta _2(z|q)&=&2q^{\frac{1}{4}}\sum_{n=0}^{\infty }
 q^{n\left(n+1\right)}\cos\left(2n+1\right)z
\nonumber \\
\theta _3(z|q)&=&1+2\sum_{n=0}^{\infty }q^{n^2}\cos 2nz
\nonumber \\
\theta _4(z|q)&=&1+2\sum_{n=0}^{\infty }\left(-1\right)^nq^{n^2}\cos 2nz,
\end{eqnarray}
and the Eisenstein series:
\begin{eqnarray}
 E_2(q)&=&1-24\sum_{n=1}^{\infty }\frac{nq^n}{1-q^n}
\nonumber \\
 E_4(q)&=&1+240\sum_{n=1}^{\infty }\frac{n^3q^n}{1-q^n}
\nonumber \\
 E_6(q)&=&1-504\sum_{n=1}^{\infty }\frac{n^5q^n}{1-q^n}.
\end{eqnarray}
The higher Eisenstein series, including $E_4$ and $E_6$, can be expressed in terms of the theta-constants:
\begin{eqnarray}
 2E_4(q^2)&=&\theta _2^8(0|q)+\theta _3^8(0|q)+\theta _4^8(0|q)
\nonumber \\
2E_6(q^2)&=&\left(\theta _4^4(0|q)-\theta _2^4(0|q)\right)\left(2\theta _3^8(0|q)+\theta _2^4(0|q)\theta _4^4(0|q)\right).
\end{eqnarray}
Together with the identity
\begin{equation}
 \theta _3^4(0|q)=\theta _2^4(0|q)+\theta _4^4(0|q),
\end{equation}
this allows one to express $\xi $, $\eta $ and $\bar{\eta }$ from (\ref{xi-eta-throughtheta}) through the Eisenstein series only:
\bea
\xi+\eta +\bar \eta&=& \frac{M^2}{4}\, E_2
\nonumber\\
\xi \eta +\xi \bar\eta +\eta\bar\eta &=& \frac{M^4}{48} \left(E_4-E_2^2\right)
\nonumber\\
\xi\eta \bar \eta &=& \frac{M^6}{1728} \left(2E_6-3 E_2E_4+E_2^3\right).
\label{utiles}
\eea
Here, as in the main text, the argument of the Eisenstein series is $-r^2$, with $r$ given by (\ref{modularr}).

The Eisenstein series $E_2$ is also not independent; it is given in terms of $E_4$ and $E_6$ by the logarithmic derivative of the modular discriminant $\Delta = 1/
1728\, (E_4^3-E_6^2)$ (the derivation being\footnote{$q=\,{\rm e}\,^{i\pi \tau }$, as usual.} $\frac{1}{2\pi i} \frac{ d}{d\tau }$). It follows that it is not a modular form: the modular transformation
has an anomalous piece,
\be
E_2(\alpha\cdot \tau)= (c\tau+d)^2 E_2(\tau ) +\frac{6c}{\pi i} (c\tau+d)\ ,\qquad \alpha\cdot \tau =\frac{a\tau+b}{c\tau+d}\ ,\ \ \alpha \in SL(2,Z)
\ee

Complete elliptic integrals $K\equiv K(m)$, $E\equiv E(m)$ are expressed through theta-constants as
\begin{eqnarray}
 m&=&\frac{\theta^4 _2}{\theta^4 _3}\label{mmmm}
 \\
 K&=&\frac{\pi }{2}\,   \theta ^2_3 \label{KKKK}
  \\  \label{EEEE}
  E&=&\frac{\pi }{6}\,\,\frac{  E_2+\theta ^4_3+\theta
   ^4_4} {\theta ^2_3}\,,
\end{eqnarray}
with
\begin{equation}
 r=-i\,{\rm e}\,^{-\frac{\pi K'}{K}},
\end{equation}
where $K'=K(1-m)$.
The incomplete elliptic integrals $E(\varphi |m)$ and $F(\varphi |m)$ are given by
\begin{eqnarray}
\sin\varphi &=&\frac{\theta _3(0)\theta _1(v)}{\theta _2(0)\theta _4(v)}
\label{ellipticsin}
\\
KE(\varphi )-EF(\varphi )&=&\frac{\pi \theta '_4(v)}{2\theta _4(v)}\,.
\label{KE-EF}
\end{eqnarray}

\section{Scaling behavior of endpoint position}\label{appendixcriticalcalculations}

In this appendix we derive (\ref{Deltathroughlambda}), (\ref{CDelta}) from the solution of the $\mathcal{N}=2^*$ matrix model in the weak-coupling phase.

First, it is convenient to express the parameters of the solution in terms of  elliptic integrals, using eqs.~(\ref{mmmm})--(\ref{KE-EF}). We have:
\begin{eqnarray}\label{xi-eta}
 \xi &=&\frac{M^2}{\pi ^2}\,KE\ ,
\nonumber \\
\eta &=&\frac{M^2}{\pi ^2}\,K(E-K)\ ,
\nonumber \\
\bar{\eta }&=&\frac{{M}^2}{\pi ^2}\,K\left[E-\left(1-m\right)K\right],
\end{eqnarray}
and the modular parameter $m$ is determined by the equation 
\begin{equation}\label{K'/K}
 \frac{K'}{K}=\frac{4\pi }{\lambda }-\frac{i}{2}\,.
\end{equation}
 The width of the eigenvalue distribution is given by
\begin{equation}\label{Pieq}
 \mu =-\frac{iM}{\pi }\left(KE(\varphi )-EF(\varphi )\right),\qquad 
 \sin^2\varphi =\frac{K-E}{mK}\,.
\end{equation}

The phase transition happens when $\xi(\lambda _c)=0 $. From (\ref{xi-eta}) we  see that at the critical point elliptic $E$ turns to zero: $E(m_c)=0$. The second equation in (\ref{Pieq}) then implies that
\begin{equation}\label{sincrit}
 \sin^2\varphi _c=\frac{1}{m_c}\,.
\end{equation}
For such $\varphi _c$, incomplete elliptic integrals can be expressed through the complete ones:
\begin{equation}\label{FEcrit}
 F(\varphi _c)=K-iK',\qquad  E(\varphi _c)=i(E'-K').
\end{equation}
Substituting these equalities into the first equation in (\ref{Pieq}), and using  Legendre's identity,
\begin{equation}\label{Legendre}
 KE'+EK'-KK'=\frac{\pi }{2}\,,
\end{equation}
along with $E=0$, we obtain that $\mu (\lambda _c)=M/2$, which demonstrates the equivalence of the two conditions for the critical coupling, $\xi =0$ and $\mu =M/2$.
 
To study the critical behavior we need the first correction to (\ref{FEcrit}). We can regard either $\varphi -\varphi _c$ or elliptic $E$ as a small parameter, since both vanish at the critical point. The two are related by the second equation in (\ref{Pieq}) and (\ref{sincrit}).
Since $\sin^2\varphi =1/m$ is a square-root branch point of $F(\varphi )$ and $E(\varphi )$, their expansion in $\varphi -\varphi _c$ contains non-analytic terms:
\begin{eqnarray}
 F(\varphi )&\simeq &K-iK'-\frac{\sqrt{\sin^2\varphi _c-\sin^2\varphi }}{\cos\varphi _c} \simeq K-iK'-\sqrt{\frac{E}{\left(m-1\right)K}} 
\nonumber \\
E(\varphi )&\simeq &i\left(E'-K'\right)+E-\frac{m\left(\sin^2\varphi _c-\sin^2\varphi \right)^{\frac{3}{2}}}{3\cos\varphi _c}\simeq i\left(E'-K'\right)+E-\sqrt{\frac{E^3}{9\left(m-1\right)K^3}}. 
\end{eqnarray}
Substituting this into (\ref{Pieq}), we get for (\ref{Deltadef}):
\begin{equation}\label{DeltaE}
 \Delta \simeq -\frac{2ME^{\frac{3}{2}}}{3\pi \sqrt{\left(1-m\right)K}}\,.
\end{equation}
Trading elliptic integrals for the parameters $\xi $ and $\eta $ with the help of (\ref{xi-eta}), we arrive at eq.~(\ref{Delta-xi}) in the main text.

Now using
\begin{eqnarray}
 &&\left(\frac{\partial E}{\partial m}\right)_{E=0}=-\frac{K}{2m}
\nonumber \\
&&\left(\frac{\partial }{\partial m}\,\,\frac{K'}{K}\right)_{E=0}
=\frac{\pi }{4m\left(m-1\right)K^2}
\end{eqnarray}
we get from (\ref{K'/K}):
\begin{equation}
 E
\simeq \frac{8\left(1-m\right)K^3}{\lambda _c^2}\left(\lambda _c-\lambda \right).
\end{equation}
Substituting this into (\ref{DeltaE}) and expressing $m$ and $K$ through the theta-constants as in (\ref{mmmm}), (\ref{KKKK}) we get eq.~(\ref{Deltathroughlambda}).

\section{Quiver models}\label{quiver:appendix}

In this appendix we  briefly comment on the large $N$ behavior of certain quiver models
which interpolate  between different theories consider in the main text, as well as ${\cal N}=4$ SYM, pure  ${\cal N}=2$ SYM
and the ${\cal N}=2$ superconformal SYM with $2N$ massless hypermultiplets in the fundamental representation.

We consider the following quivers:

\noindent A) $SU(N)_1\times SU(N)_2$ with one 
bi-fundamental hyper and
$N$ fundamental hypers for each of the $SU(N)$'s.
The localization partition function of this superconformal theory is
\bea
Z_{\rm A} &=&  \int d^{N-1}a\ d^{N-1}b \, \frac{\prod_{i<j}^{}\left(a_i-a_j\right)^2 \left(b_i-b_j\right)^2 H^2(a_i-a_j)H^2(b_i-b_j)}{\prod_{i,j}H(a_i-b_j) \prod_{i} H^N(a_i) H^N(b_i)  }
\,{\rm e}^{- 8\pi^2 N\sum\limits_{i} \frac{ a_i ^2} {\lambda _1}+  \frac{ b_i ^2} {\lambda _2}}
\nonumber\\
&\times & \left|{\cal Z }_{\rm inst}^{\rm A}(a,b;g_1^2,g_2^2)\right|^2.
\label{modA}
\eea
 If one of the two couplings is set to zero (for instance, $\lambda _2$), the $b_i$ variables are frozen at $b_i=0$, and the theory  becomes equivalent to  ${\cal N}=2$ $SU(N)$ SCYM with $2N$ flavors. 

\noindent B) $SU(N)_1\times SU(N)_2$ with 2  bi-fundamental hypers. The localization formula gives
\bea
Z_{\rm B} &=&  \int d^{N-1}a\ d^{N-1}b \, \frac{\prod_{i<j}^{}\left(a_i-a_j\right)^2 \left(b_i-b_j\right)^2 H^2(a_i-a_j)H^2(b_i-b_j)}{\prod_{i,j}H^2(a_i-b_j)  }
\,{\rm e}^{- 8\pi^2 N\sum\limits_{i} \frac{ a_i ^2} {\lambda _1}+  \frac{ b_i ^2} {\lambda _2}}
\nonumber\\
&\times &\left|{\cal Z }_{\rm inst}^{\rm B}(a,b;g_1^2,g_2^2)\right|^2.
\label{modeB}
\eea
This theory is also superconformal. Decoupling one of the gauge groups again gives  ${\cal N}=2$ SCYM with $2N$ flavors. 

\noindent B$^*$) The massive deformation of the B model:
\bea
Z_{\rm B^*} &=&  \int d^{N-1}a\ d^{N-1}b \, \frac{\prod_{i<j}^{}\left(a_i-a_j\right)^2 \left(b_i-b_j\right)^2 H^2(a_i-a_j)H^2(b_i-b_j)}{\prod_{i,j}H(a_i-b_j+M) H(a_i-b_j-M)   }
\,{\rm e}^{- 8\pi^2 N\sum\limits_{i} \frac{ a_i ^2} {\lambda _1}+  \frac{ b_i ^2} {\lambda _2}}
\nonumber\\
&\times &\left|{\cal Z }_{\rm inst}^{\rm B^*}(a,b;g_1^2,g_2^2)\right|^2\ .
\label{massB}
\eea
The theory with $\lambda_1=\lambda_2$ should have similar  dynamics as ${\cal N}=2^*$ $SU(N)$ SYM. The hypermultiplets decouple in the IR and the theory flows to two copies of pure $\mathcal{N}=2$ SYM.

\subsection{ Model A}

After introducing two densities for the $a_i$ and $b_i$ eigenvalues,
 the saddle-point equations become
\be
\strokedint_{-\mu_1}^{\mu_1} dy \rho_1(y) \left(\frac{1}{x-y} -\mathcal{K}(x-y)\right) + \frac{1}{2} \strokedint_{-\mu_2}^{\mu_2} d\hat x \rho_2(\hat x)\mathcal{K}( x- \hat x)
= \frac{8\pi^2}{\lambda_1}\,x -\frac{\mathcal{K}(x)}{2}\ ,
\label{dsss}
\ee
\be
\strokedint_{-\mu_2}^{\mu_2} d\hat y \rho_2(\hat y) \left(\frac{1}{\hat x-\hat y} -\mathcal{K}(\hat x-\hat y)\right) +\frac{1}{2}\strokedint_{-\mu_1}^{\mu_1} dx \rho_1( x)\mathcal{K}( \hat x- x)
= \frac{8\pi^2}{\lambda_2}\, \hat x -\frac{\mathcal{K}(\hat x)}{2}\ .
\label{dos}
\ee
When $\lambda _2\ll 1$, the $\rho _2$ density is peaked at zero, while $\rho _1$ satisfies the equations for  ${\cal N}=2$ SCYM, as expected.

Another simplifying case is $\lambda_1=\lambda_2$, when the model possesses a $\mathbbm{Z}_2$ symmetry. The solution with unbroken $\mathbbm{Z}_2$ has $\rho_1=\rho_2$, and we are left with the equation
\be
\strokedint_{-\mu_1}^{\mu_1} dy \rho_1(y) \left(\frac{2}{x-y} -\mathcal{K}(x-y)\right) 
= \frac{16\pi^2}{\lambda_1}\,x -\mathcal{K}(x).
\label{dss}
\ee
This equation  is similar but not equivalent to the one for ${\cal N}=2$ SCFT with $2N$ massless hypermultiplets.

\subsection{ Model B}

The saddle-point equations are 
\be
\strokedint_{-\mu_1}^{\mu_1} dy \rho_1(y) \left(\frac{1}{x-y} -\mathcal{K}(x-y)\right) +\strokedint_{-\mu_2}^{\mu_2} d\hat x \rho_2(\hat x)\mathcal{K}( x- \hat x)
= \frac{8\pi^2}{\lambda_1}\,x \ ,
\nonumber
\ee
\be
\strokedint_{-\mu_2}^{\mu_2} d\hat y \rho_2(\hat y) \left(\frac{1}{\hat x-\hat y} -\mathcal{K}(\hat x-\hat y)\right) +\strokedint_{-\mu_1}^{\mu_1} dx \rho_1( x)\mathcal{K}( \hat x- x)
= \frac{8\pi^2}{\lambda_2}\, \hat x\ .
\label{cuatro}
\ee
Interestingly, at the symmetric point $\lambda_1=\lambda_2\equiv \lambda $ the complicated interaction terms cancel, and the saddle-point equations reduce to those for the Gaussian matrix model. The solution is   
\be
\rho_1( x)=\rho _2(x)=\frac{2}{\pi\mu^2} \sqrt{\mu^2- x^2}\ ,\ \qquad \mu=\frac{\sqrt{\lambda}}{2\pi}\ .
\nonumber
\ee

\subsection{Massive B$^*$ model}

For the massive case, the B$^*$ model,  the saddle-point equations read
\be
\label{kol}
\strokedint_{-\mu_1}^{\mu_1} dy \rho_1(y) \left(\frac{1}{x-y} -\mathcal{K}(x-y)\right) + \frac{1}{2}\strokedint_{-\mu_2}^{\mu_2} d\hat x \rho_2(\hat x)(\mathcal{K}( x- \hat x+M)+\mathcal{K}( x- \hat x-M))
= \frac{8\pi^2}{\lambda_1}\,x \ ,
\nonumber
\ee
\be
\label{kolt}
\strokedint_{-\mu_2}^{\mu_2} d\hat y \rho_2(\hat y) \left(\frac{1}{\hat x-\hat y} -\mathcal{K}(\hat x-\hat y)\right) + \frac{1}{2}\strokedint_{-\mu_1}^{\mu_1} dx \rho_1( x)(\mathcal{K}( \hat x- x+M)+\mathcal{K}( \hat x- x-M))
= \frac{8\pi^2}{\lambda_2}\, \hat x\ .
\nonumber
\ee
If $\lambda _2\rightarrow 0$, the second gauge group decouples leaving behind  $\mathcal{N}=2$ SYM with $2N$ massive hypermultiplets. On the other hand, if $\lambda _1=\lambda _2$, the symmetric solution satisfies the saddle-point equation for $\mathcal{N}=2^*$ SYM (\ref{nnstar}).
Thus the $B^*$ model  interpolates between the two massive models discussed 
in the main text.

\bibliographystyle{nb}

\end{document}